\documentclass[
 reprint,
 floatfix,          
 amsmath,amssymb,
 aps,
 prd,
 superscriptaddress,
 nofootinbib,
]{revtex4-2}

\usepackage{graphicx}
\usepackage{dcolumn}
\usepackage{bm}
\usepackage{xcolor}
\usepackage{amsmath}
\usepackage{multirow}
\usepackage{hyperref}
\usepackage[utf8]{inputenc}

\newcommand{\LCDM}{\ensuremath{\Lambda\mathrm{CDM}}}
\newcommand{\wowa}{\ensuremath{w_0w_a\mathrm{CDM}}}
\newcommand{\LBDM}{\ensuremath{\Lambda\mathrm{BDM}}}
\newcommand{\wowaBDM}{\ensuremath{w_0w_a\mathrm{BDM}}}

\newcommand{\ac}{\ensuremath{a_c}}
\newcommand{\Lc}{\ensuremath{\Lambda_c}}
\newcommand{\Neff}{\ensuremath{N_{\mathrm{eff}}}}
\newcommand{\Next}{\ensuremath{N_{\mathrm{ext}}}}
\newcommand{\vc}{\ensuremath{v_c}}
\newcommand{\fbde}{\ensuremath{f_{\mathrm{BDE}}}}
\newcommand{\Ode}{\ensuremath{\Omega_{\mathrm{DE}}}}
\newcommand{\Om}{\ensuremath{\Omega_{m}}}
\newcommand{\sig}{\ensuremath{\sigma_8}}
\newcommand{\Seight}{\ensuremath{S_8}}
\newcommand{\Hub}{\ensuremath{H_0}}
\newcommand{\khalf}{\ensuremath{k_{1/2}}}
\newcommand{\rhoDG}{\ensuremath{\rho_{\rm DG}}}
\newcommand{\ca}{\ensuremath{c_a^2}}
\newcommand{\ceff}{\ensuremath{c_{\rm eff}^2}}
\newcommand{\cvis}{\ensuremath{c_{\rm vis}^2}}

\begin{document}

\title{The unified Bound Dark Energy--Bound Dark Matter model:
\texorpdfstring{\\}{ }
theory and constraints from DESI DR2}

\author{Jose Agustin Lozano Torres}
\email{lozano@estudiantes.fisica.unam.mx}
\affiliation{Instituto de F\'isica, Universidad Nacional Aut\'onoma de
M\'exico, Ciudad de M\'exico, 04510, Mexico}

\author{Axel de la Macorra}
\affiliation{Instituto de F\'isica, Universidad Nacional Aut\'onoma de
M\'exico, Ciudad de M\'exico, 04510, Mexico}

\date{\today}

\begin{abstract}
The unified Bound Dark Energy--Bound Dark Matter (BDE-BDM) model derives
both the dark energy and the whole of the dark matter from a single
supersymmetric $SU(3)$ gauge group with $N_f=6$ flavors. Gauge-coupling
unification predicts the scale at which it condenses,
$\Lc=34^{+16}_{-11}$~eV, before any cosmological data are used, so the
dark-energy sector adds no parameter to those of \LCDM\ and the dark-matter
sector adds only one. Fitting DESI DR2 BAO, Planck CMB, Planck+ACT DR6 CMB lensing,
and three supernova compilations (PantheonPlus, Union3.1, DES-Dovekie)
gives, for Union3.1, $\Lc=43.67\pm0.13$~eV, inside the predicted range: a high-energy
prediction met by a late-time measurement. The equation of state evolves
to $w_0=-0.9262\pm0.0005$ and never crosses the phantom divide. Relative
to \LCDM\ the model lowers the DESI BAO $\chi^2$ by $2.95$--$3.67$ and
the supernova $\chi^2$ by $2.53$--$4.07$, recovering $57\%$--$71\%$ of
the total reduction that \wowa\ obtains by fitting two dark-energy
parameters to the same data, and reaching the lower BAO $\chi^2$ of the
two models for two of the three compilations. Its $\Delta\mathrm{DIC}$
against \LCDM\ is strong for every compilation, $-8.28$ to $-7.47$,
varying by $0.81$ where that of \wowa\ varies by $4.8$. Because the
dark-energy sector adds no parameter to those of \LCDM, the likelihood gain survives
the Occam penalty: $\ln B=+4.09$, $+4.12$, and $+4.83$ against \LCDM,
where \wowa\ gives $-2.11$, $-0.26$, and $-0.56$. Among six models pairing
a cosmological constant, the \wowa\ parametrization or bound dark energy
with cold or bound dark matter, BDE-BDM has the largest Bayesian evidence.
\end{abstract}

\maketitle

\section{Introduction}
\label{sec:intro}

For over two decades, the accelerated expansion of the Universe has primarily been attributed to a cosmological constant, $\Lambda$. The \LCDM\ (Lambda Cold Dark Matter) model, built upon this assumption, has consistently aligned with significant observational evidence. However, it lacks a robust theoretical foundation: $\Lambda$ remains an unknown parameter with no specific properties or established physical origin, and its measured value is notably lower than predictions from quantum field theory.

Recent analyses, particularly those using baryon acoustic oscillation (BAO) data from the Dark Energy Spectroscopic Instrument (DESI) combined with Cosmic Microwave Background (CMB) and supernova observations, hint at a preference for a time-varying dark energy density~\cite{DESI2024VI,DESI2025II}. Fitting these data with the \wowa\ parameterization $w(a)=w_0+w_a(1-a)$~\cite{Chevallier2001,Linder2003} yields $w_0 > -1$ and $w_a < 0$, implying an equation of state that evolves with time and crosses $w = -1$ at low redshift. The significance of this deviation from \LCDM\ ranges between $2.8\sigma$ and $4.2\sigma$ depending on the supernova sample~\cite{DESI2025II}, the upper value falling to $3.2\sigma$ with the recalibrated DES-Dovekie sample~\cite{DESDovekie}, and remains near the $3\sigma$ threshold even when these are excluded from the analysis. Additionally, other reconstruction methods and data partitions corroborate this trend~\cite{Lodha2025,GiareDDE2024,Ishak2025}.

Yet, while the \wowa\ model provides a fitting function, it offers no fundamental explanation; its parameters are free and not derived from any underlying theory. Here, we present the unified Bound Dark Energy–Bound Dark Matter (BDE-BDM) model, which addresses these questions. Within this framework, dark energy and the whole of the dark matter originate from a single supersymmetric $SU(N_c)$ gauge group that remains decoupled from the Standard Model~\cite{Macorra2005, Macorra2018, Almaraz2019, Macorra2010, Mastache2020, MacorraLozano2025}. The dynamics are analogous to those of quantum chromodynamics (QCD): in the early Universe, as it cools, the coupling of these dark particles evolves toward strong coupling, confining elementary fields into neutral composite states—similar to the formation of hadrons from quarks and gluons.

In cosmological terms, the lightest mesons exhibit behavior akin to a canonical scalar field, with their gauge-generated potential driving late-time cosmic acceleration—this constitutes the Bound Dark Energy (BDE)~\cite{MacorraJHEP2003,Macorra2018,Almaraz2019}. Meanwhile, heavier composite states acquire mass through binding dynamics, forming a dark matter component (Bound Dark Matter, BDM) at the transition, characterized by a velocity scale \vc~\cite{Macorra2010,Mastache2020}. Given their relativistic nature at formation, BDM particles preferentially stream out of small-scale perturbations, leading to a suppression of power below a certain wavenumber. This characteristic was proposed as a way to address the discrepancy between cold dark matter simulations and observed satellite galaxies around a Milky Way-like halo~\cite{Macorra2010, Mastache2020}; it motivates the present construction, although the scales involved lie beyond the reach of the data used here and we make no claim about satellite abundances. Thus, a single condensation event simultaneously accounts for the late-time acceleration and imposes a specific prediction regarding small-scale structure formation.

The BDE-BDM model's predictive framework is given by the non-arbitrary nature of its scale. By deriving $\Lc=34^{+16}_{-11}$ eV from a downward-running dark coupling stemming from unification, and ensuring consistency with the observed radiation temperature, the link between \Lc\ and the condensation scale factor \ac\ is established. Consequently, once \Lc\ is specified the dark-energy sector has no further freedom: the present dark-energy density, the evolution of $w$ with redshift, and the timing of the transition are direct outputs. The dark matter sector introduces one additional parameter, \vc, which quantifies the extent of power suppression due to free streaming. This contrasts with warm dark matter scenarios; here, the non-relativistic transition epoch is inherently tied to the same condensation process driving the dark energy.

This paper offers a comprehensive explanation of the model, detailing both background dynamics and linear perturbation theory, including the energy-conserving distribution of the dark-group density between the two bound components during condensation—a topic only briefly addressed in previous work. It also rigorously compares the model against current data, evaluating its performance relative to \LCDM\ and \wowa.

\section{The unified BDE-BDM model}
\label{sec:model}

\textit{The dark gauge sector.}---The dark sector is a supersymmetric
$SU(N_c)$ gauge theory with $N_c=3$ colors and $N_f=6$ flavors of
massless dark quarks, interacting with the Standard Model only
gravitationally~\cite{Burgess1997,MacorraJHEP2003,Macorra2005}. Above
the condensation scale, the theory is asymptotically free and its
elementary particles---dark gluons, dark quarks (namely dark plasma hereafter) and their
superpartners---are massless and relativistic, contributing
\begin{equation}
  g_{\rm DG}
  = \left(1+\tfrac{7}{8}\right)
    \left[\,2\,(N_c^2-1) + 2\,N_c N_f\,\right]
  = 97.5
  \label{eq:gDG}
\end{equation}
effective relativistic degrees of
freedom~\cite{MacorraJHEP2003,Almaraz2019}, the factor
$(1+\tfrac78)$ accounting for the equal boson and fermion content of
the super symmetric multiplets. At one loop the running is controlled by
$b_0=3N_c-N_f=3$, which is positive, so the coupling grows toward low
energies. If the dark and Standard Model couplings unify at
$\Lambda_{\rm gut}=(1.05\pm0.07)\times10^{16}$~GeV with
$g_{\rm gut}^2=4\pi/(25.83\pm0.16)$~\cite{Bourilkov2015,PDG2024}, the
dark coupling becomes strong, and the sector condenses, at
\begin{equation}
  \Lc = \Lambda_{\rm gut}\,
        e^{-8\pi^2/(b_0\, g_{\rm gut}^2)}
      = 34^{+16}_{-11}~\mathrm{eV},
  \label{eq:Lc}
\end{equation}
The uncertainty is dominated by that of the unification coupling
through the exponential, similarly as the QCD scale follows from the running of the strong coupling.

Before condensation, the dark group is a relativistic (dark) plasma, at a temperature below the photon temperature because the
two sectors decouple at unification and the Standard Model subsequently
reheats its own bath as heavy species annihilate. Entropy conservation
in each sector, with the plasma density matched to the present photon
density $\rho_{\gamma0}=(\pi^2/15)T_0^4$ at $T_0=2.7255$~K, gives the
second fundamental relation~\cite{Macorra2018,Almaraz2019},
\begin{equation}
\begin{split}
  \frac{\ac\Lc}{\mathrm{eV}}
  &= \left(\frac{\rho_{r0}}{3\,\mathrm{eV}^4}\,
          \frac{g_{\rm DG}}{g_r}\right)^{1/4}
    \left(\frac{4}{11}\,
          \frac{g_{\rm SM}^{\nu{\rm dec}}}{g_{\rm SM}^{\rm gut}}
    \right)^{1/3}\\
  &= 1.0939\times10^{-4},
\end{split}
  \label{eq:acLc}
\end{equation}
with $g_{\rm SM}^{\rm gut}=228.75$ the Standard Model degrees of freedom
at unification, $g_{\rm SM}^{\nu{\rm dec}}=10.75$ their value at neutrino
decoupling, and $g_r=2+\tfrac78(2)N_\nu(T_\nu/T_\gamma)^4=3.383$ with
$N_\nu=3.044$ the relativistic degrees of freedom of the present
radiation density $\rho_{r0}=(\pi^2/30)\,g_r T_0^4$. Because $\rho_{r0}/g_r=\rho_{\gamma0}/2$ identically, the
product $\ac\Lc$ does not depend on the effective number of neutrino
species assumed for the Standard Model bath. Together with
Eq.~\eqref{eq:Lc} it places the transition deep in the radiation era:
for \Lc\ near its predicted value, $\ac\simeq2.5\times10^{-6}$ and
$T_\gamma(\ac)\simeq94$~eV, well after big-bang nucleosynthesis (BBN)
and well before matter--radiation equality.\footnote{We quote
$\ac\Lc=1.0939\times10^{-4}$~eV, the value given in Eq.~(15) and in the
conclusions of Ref.~\cite{Almaraz2019}; the abstract of that work quotes
$1.0934\times10^{-4}$~eV. Recomputing the constant from the gauge inputs,
either through \Next\ or through the density ratio of its Eq.~(12),
gives $1.09393\times10^{-4}$~eV.} The normalization
$\rhoDG(\ac)=3\Lc^4$ defines the transition: condensation occurs when
the plasma density reaches the value set by the condensation scale
itself, and before it
\begin{equation}
  \rhoDG(a) \;=\; 3\,\Lc^4\left(\frac{\ac}{a}\right)^{4},
  \qquad a\le\ac .
  \label{eq:rhoDG}
\end{equation}
As in the BDE and BDM
analyses~\cite{MacorraLozano2025,Mastache2020}, the subscript $c$ denotes
quantities evaluated at \ac, the dark plasma is described for $a\le\ac$
and the bound components for $a\ge\ac$, and relations at \ac\ connect the
two descriptions.
Equivalently, $\rhoDG(\ac)=3\Lc^4$ fixes the temperature of the dark
plasma at condensation,
$T_c=\big[90/(\pi^2 g_{\rm DG})\big]^{1/4}\Lc\simeq0.55\,\Lc$,
slightly below the scale at which the coupling diverges, as it should
be. In the standalone bound-dark-energy model with cold dark matter
(BDE-CDM), this same normalization
was equivalent to the choice $w_{\rm BDE}(\ac)=1/3$, since
$\rho_{\rm BDE}(\ac)=2\Lc^4/[1-w_{\rm BDE}(\ac)]$ there~\cite{Almaraz2019};
here it is retained as a thermodynamic property of the dark plasma,
independent of how the condensation energy is subsequently divided, and
$w_{\rm BDE}(\ac)$ follows from the partition instead, as
Eq.~\eqref{eq:phidotc} below makes explicit.

The dark plasma adds to the pre-condensation expansion rate an amount
conveniently written as an extra effective number of neutrino
species~\cite{Almaraz2019},
\begin{equation}
  \Next
  = \frac{4}{7}\,g_{\rm DG}
    \left(\frac{g_{\rm SM}^{\nu{\rm dec}}}{g_{\rm SM}^{\rm gut}}
    \right)^{4/3}
  = 0.945,
  \label{eq:Next}
\end{equation}
amounting at the transition to a fraction
\begin{equation}
  \frac{\rhoDG(\ac)}{\rho_r(\ac)}
  = \frac{g_{\rm DG}}{g_r}
    \left(\frac{4}{11}\,
          \frac{g_{\rm SM}^{\nu{\rm dec}}}{g_{\rm SM}^{\rm gut}}
    \right)^{4/3}
  = 0.1268
  \label{eq:rhoDGoverrhor}
\end{equation}
of the standard radiation density, a little over a tenth of the total
energy density at that epoch; in evaluating the latter, the matter term
contains baryons alone, since the dark matter of this model is itself
part of the dark group before condensation. Two features distinguish
Eq.~\eqref{eq:Next} from a generic $\Delta\Neff$. Its value is not free,
following from the group content through
Eqs.~\eqref{eq:gDG}--\eqref{eq:acLc}; and it is transient, since after
\ac\ the dark-group energy is converted into the two bound components,
whose combined density immediately dilutes faster than radiation. The CMB therefore
sees an essentially standard radiation content around recombination,
while BBN, part of which occurs while the dark plasma is still present, is
affected in a time-dependent way (Sec.~\ref{sec:discussion}).

\textit{Bound dark energy.}---At the transition the lightest bound state
is a dark meson, described by a canonically normalized scalar field
$\phi$ whose effective potential is generated non-perturbatively by the
gauge dynamics and takes the Affleck--Dine--Seiberg inverse-power-law
form~\cite{Affleck1984,Burgess1997,MacorraJHEP2003,Almaraz2019}
\begin{equation}
  V(\phi) = \Lc^{\,4+n}\,\phi^{-n},
  \qquad
  n = 2\left(1+\frac{2}{N_c-N_f}\right) = \frac{2}{3},
  \label{eq:Vphi}
\end{equation}
where both the exponent and the overall scale follow from the group
content and Eq.~\eqref{eq:Lc}. The associated mass,
$m_\phi^2=(10/9)(\Lc/\phi)^{8/3}\Lc^2$, is protected by supersymmetry and
by the global symmetries of the condensate, and radiative corrections to
Eq.~\eqref{eq:Vphi} are suppressed by $(\Lc/\phi)^2/(32\pi^2)$ and
negligible once the field rolls away from the
transition~\cite{Almaraz2019}. No parameter of the dark-energy sector is
chosen by hand, which distinguishes the model from quintessence with the
same functional form of $V(\phi)$, where the scale multiplying the
potential is adjusted to reproduce the observed acceleration.

The field starts at \ac\ with the energy inherited from the condensing
fraction of the dark group plasma. Writing \fbde\ for the fraction that condenses
into the meson, the remainder forming the bound dark matter, the initial
conditions are
\begin{equation}
\begin{split}
  &\phi_c = \Lc, \qquad V(\phi_c) = \Lc^4, \\
  &\rho_{\rm BDE}(\ac) = \fbde\,\rhoDG(\ac) = 3\,\fbde\,\Lc^4 ,
\end{split}
  \label{eq:rhobdeic}
\end{equation}
The value of $\phi_c$ is the symmetry-breaking scale of the dark
sector itself~\cite{MacorraJHEP2003,Almaraz2019}, fixed by the gauge
dynamics and therefore independent of how the condensation energy is
subsequently divided. Since $V(\phi_c)$ is fixed along with it, the
partition determines the kinetic term,
$\tfrac12\dot\phi_c^2=(3\fbde-1)\Lc^4$, where overdots denote derivatives
with respect to cosmic time, and hence
\begin{equation}
  \dot\phi_c = \sqrt{6\fbde-2}\;\Lc^2 ,
  \qquad
  w_{\rm BDE}(\ac) = \frac{3\fbde-2}{3\fbde} ,
  \label{eq:phidotc}
\end{equation}
so that the equation of state at the transition is an output of the
partition rather than an input. Energy is conserved exactly across the
transition; the pressure is not, the difference being the latent heat
released as the plasma confines and the trace anomaly changes, which
leaves $\rho$ continuous and $\dot\rho=-3H(\rho+P)$, with $H=\dot a/a$,
integrable through
\ac. For \fbde\ close to unity, the field emerges with an equation of
state just below the plasma value $1/3$, and in the limit $\fbde\to1$
Eq.~\eqref{eq:phidotc} reduces to $\dot\phi_c=2\Lc^2$ and
$\rho_{\rm BDE}(\ac)=3\Lc^4$, the initial conditions of
BDE-CDM~\cite{Macorra2018,Almaraz2019}. Positivity of the kinetic term
requires
\begin{equation}
  \tfrac13 \le \fbde \le 1 ,
  \label{eq:fbdebound}
\end{equation}
a hard physical bound rather than a prior choice: with $V(\phi_c)=\Lc^4$
fixed, the meson cannot inherit less than $\Lc^4$ however slowly the
field moves, so the bound dark matter cannot take more than two thirds
of the condensation energy. Thereafter the field obeys the
Klein--Gordon equation in the expanding background,
\begin{equation}
  \ddot\phi + 3H\dot\phi + \frac{dV}{d\phi} = 0 ,
  \label{eq:kg}
\end{equation}
with $\rho_{\rm BDE}=\tfrac12\dot\phi^2+V$ and
$P_{\rm BDE}=\tfrac12\dot\phi^2-V$.

Its history has three stages, shown in the left panel of
Fig.~\ref{fig:background}.
Immediately after \ac\ the kinetic term dominates, and the field enters a
stiff phase with $w\simeq1$ and $\rho_{\rm BDE}\propto a^{-6}$, so the
dark-energy density, initially a tenth of the total, becomes negligible
within a few $e$-folds; this same dilution removes the extra radiation
of Eq.~\eqref{eq:Next} and keeps the dark energy negligible through
radiation and matter domination without any adjustment of the initial
data. The field then freezes by Hubble friction, its density staying
nearly constant and far below the background with $w\simeq-1$. Finally,
when the background density falls to the level of the frozen field, at
redshifts of order unity, the potential takes over and the field thaws,
$w$ rising from $-1$ toward a present value that stays above the phantom
divide, so the instabilities associated with $w<-1$ never arise. This
entire late-time history is a function of \Lc\ alone, and the expansion
history it generates is of the thawing, non-phantom kind the DESI data
prefer~\cite{DESI2025II,Lodha2025} --- a consequence of
Eqs.~\eqref{eq:Lc}--\eqref{eq:Vphi} rather than of a fit.

\textit{Bound dark matter.}---The heavier bound states formed at the
transition acquire their mass non-perturbatively from the binding, as
most of the proton mass arises from QCD rather than from the quark
masses~\cite{Macorra2010,Mastache2020}. In the original formulation, the
particles are massless and relativistic above a phase-transition energy
density $\rho_{\rm BDM}(\ac)$ and
non-relativistic below it, so that the free-streaming length
$\lambda_{\rm fs}\propto\rho_{\rm BDM}(\ac)^{-1/3}$ and the
free-streaming mass $M_{\rm fs}\propto\rho_{\rm BDM}(\ac)^{-1}$ are both
controlled by that single scale, which structure-formation arguments
place at $\rho_{\rm BDM}(\ac)^{1/4}=\mathcal{O}(10\text{--}100)$~eV~\cite{Macorra2010}.
The essential point of the unified model is that this scale is not
independent. The massive states inherit the complement of the fraction
that condenses into the meson, so that
\begin{equation}
\begin{aligned}
  \rho_{\rm BDM}(\ac) &= 3\,(1-\fbde)\,\Lc^4 , \\[2pt]
  \lambda_{\rm fs} &\propto \big[3(1-\fbde)\big]^{-1/3}\,\Lc^{-4/3},\\[2pt]
  M_{\rm fs} &\propto \big[3(1-\fbde)\big]^{-1}\,\Lc^{-4},
\end{aligned}
  \label{eq:rhobdmc}
\end{equation}
with \fbde\ itself determined below by Eq.~\eqref{eq:fbde}: the
small-scale phenomenology of the dark matter and the late-time
acceleration are therefore controlled by the same \Lc\ of
Eq.~\eqref{eq:Lc}, fixed by gauge-coupling unification, together with the
partition it undergoes at condensation. Expressed as an equivalent
energy scale for comparison with the earlier literature,
$\rho_{\rm BDM}(\ac)^{1/4}=[3(1-\fbde)]^{1/4}\Lc\simeq0.66\,\Lc$, which
for the parameters preferred by the data lies inside the range quoted
above while remaining a derived quantity rather than a free one.

Each particle is relativistic before the transition and emerges from it with velocity \vc; thereafter
its momentum redshifts freely, $p\propto1/a$, so that
\begin{equation}
  v(a) = \frac{x}{\sqrt{1+x^2}},
  \qquad
  x \equiv \gamma_c\,\vc\,\frac{\ac}{a},
  \qquad
  \gamma_c=\frac{1}{\sqrt{1-\vc^2}} .
  \label{eq:va}
\end{equation}
The corresponding equation of state, $w=v^2/3$ for free particles,
interpolates between radiation and cold matter,
\begin{equation}
  w_{\rm BDM}(a) =
  \frac{1}{3}
  \left[\,1+\frac{1-\vc^2}{\vc^2}
        \left(\frac{a}{\ac}\right)^{2}\right]^{-1}
  \quad (a\ge\ac),
  \label{eq:wbdm}
\end{equation}
with $w_{\rm BDM}(\ac)=\vc^2/3$, while $w_{\rm BDM}=1/3$ for $a<\ac$;
the two branches are joined numerically by a
smooth step of width $\ac/8$, a smoothing the earlier analysis also
required~\cite{Mastache2020}. Today
$w_{\rm BDM}\sim10^{-12}$ for the parameters of interest. Equation~\eqref{eq:wbdm} satisfies
$d\ln w/d\ln a=-2(1-3w)$, so for $a\ge\ac$ energy conservation
integrates in closed form to
\begin{equation}
\begin{split}
  \rho_{\rm BDM}(a)
  &= \rho_{\rm BDM,0}\;a^{-4}\,
    \sqrt{\frac{w_{\rm BDM}(1)}{w_{\rm BDM}(a)}}
   = \rho_{\rm BDM,0}\;a^{-3}\,\frac{\gamma(a)}{\gamma(1)} ,\\
  \gamma(a) &\equiv \sqrt{1+\gamma_c^2\vc^2\left(\frac{\ac}{a}\right)^{2}} ,
\end{split}
  \label{eq:rhobdm}
\end{equation}
which tends to pressureless matter well after the transition; the
residual factor $1/\gamma_c$ records the kinetic energy lost by
particles that were relativistic when they formed. Before the
transition the component is the share $1-\fbde$ of the plasma destined to
form the massive states, $\rho_{\rm BDM}=(1-\fbde)\,\rhoDG$, and dilutes
as radiation from the value Eq.~\eqref{eq:rhobdm} takes at \ac,
\begin{equation}
\begin{aligned}
  \rho_{\rm BDM}(a) &= \rho_{\rm BDM}(\ac)\left(\frac{\ac}{a}\right)^{4} ,\\[2pt]
  \rho_{\rm BDM}(\ac) &= \gamma_c\,\rho_{\rm BDM,0}\,\ac^{-3} ,
\end{aligned}
  \label{eq:rhobdmpre}
\end{equation}
for $a<\ac$, where $\gamma(1)=1$ to within
$\gamma_c^2\vc^2\ac^2\lesssim10^{-11}$ has been used; the density is
therefore continuous at the transition, and the dark-group energy passes
through it without loss. The squared adiabatic sound speed
$\ca\equiv w-\dot w/[3H(1+w)]$ is
\begin{equation}
  \ca = \frac{w\,(5-3w)}{3\,(1+w)} ,
  \label{eq:ca2}
\end{equation}
exactly, outside the matching window, interpolating between $\ca=1/3$ in
the relativistic limit and $\ca\to\tfrac53w\to0$ in the cold limit. Two
limits are useful: for $\vc\to0$ the component is cold at the
transition, recovering Ref.~\cite{Macorra2010} and being
indistinguishable from cold dark matter on the scales resolved here; for
\vc\ approaching the upper limit $1/\sqrt2$, the value at which the
momentum equals the mass, it stays semi-relativistic for several
$e$-folds after \ac\ and behaves as a warm species. Larger velocities
would leave the bound states relativistic after the transition and are
excluded by construction~\cite{Mastache2020}, which fixes the prior of
Sec.~\ref{sec:data}. The velocity \vc\ is the single free parameter of the
dark-matter sector.

The unified model requires the dark-group energy at the transition to be shared between the two
bound components without loss,
$\rhoDG(\ac)=\rho_{\rm BDE}(\ac)+\rho_{\rm BDM}(\ac)$, which defines
the condensing fraction
\begin{equation}
  \fbde = 1-\frac{\rho_{\rm BDM}(\ac)}{\rhoDG(\ac)}
        = 1-\frac{\gamma_c\,\rho_{\rm BDM,0}\,\ac^{-3}}{3\Lc^4} ,
  \label{eq:fbde}
\end{equation}
where the second equality uses the kinematics of Eq.~\eqref{eq:va} to
relate the BDM density at the transition to its present value. Given the
present dark-matter density $\omega_{\rm bdm}$ and the velocity \vc, the
fraction \fbde\ is determined and is not an independent parameter: of
the $\Omega_{\rm DG}(\ac)$ carried by the dark group, a share \fbde\
goes to the meson and the rest to the massive states, which gain on
radiation thereafter until they become the dark matter we observe. One
consequence deserves note. Since Eq.~\eqref{eq:fbde} moves energy
between the two products without creating or destroying any, the value
of \Next\ before the transition is untouched by $(\fbde,\vc)$: what the
sector was is fixed by the gauge content, and only what it turns into
depends on the split.

\section{Linear perturbations}
\label{sec:perturbations}

We work in the synchronous gauge with the conventions of
Ref.~\cite{MaBertschinger1995}, as in the BDE and BDM
analyses~\cite{Almaraz2019,Mastache2020}: primes denote derivatives with
respect to the conformal time $\tau$, $\mathcal{H}=a'/a$, and $h$ is the
trace of the metric perturbation (not to be confused with the
dimensionless Hubble constant). Three dark-sector components require
perturbation equations: the pre-condensation plasma, the BDE field, and
the BDM fluid.

\textit{Plasma and scalar field.}---The strongly self-coupled plasma
admits no exact first-principles treatment short of a dedicated
Boltzmann hierarchy. Two limiting descriptions bracket the physics: a
perfect relativistic fluid without anisotropic stress, as for the
tightly coupled photon--baryon plasma, and a collisionless relativistic
gas, free streaming, as for massless neutrinos. The internal equilibrium
invoked in Sec.~\ref{sec:model} points to the first, since every
dark-group state carries dark color, and the sector is strongly coupled
throughout the relevant history; we nevertheless adopt the free-streaming limit, so
that the $\fbde\to1$ limit of the unified model reproduces the published
bound-dark-energy analysis~\cite{Almaraz2019} without a change of
prescription. Slaving the dark-group perturbations to the
massless-neutrino multipoles,
\begin{equation}
  \delta_{\rm DG}=\delta_\nu,\quad
  \theta_{\rm DG}=\theta_\nu,\quad
  \sigma_{\rm DG}=\sigma_\nu \qquad (a<\ac),
  \label{eq:slaved}
\end{equation}
including the anisotropic-stress contribution to the metric sources.
Consistency of the superhorizon adiabatic solution then requires the
dark-group share to be counted in the free-streaming fraction entering
the adiabatic initial conditions,
\begin{equation}
  R_\nu \;=\;
  \frac{\bar\rho_\nu+\rhoDG}{\bar\rho_\nu+\rhoDG+\bar\rho_\gamma}
  \qquad (a<\ac),
  \label{eq:Rnu}
\end{equation}
which with $\Next=0.945$ raises $R_\nu$ from its \LCDM\ value $0.409$ to
$0.475$ and suppresses the initial matter perturbations by the factor
$0.984$ found in Ref.~\cite{Almaraz2019}. In the opposite limit, the dark group is a perfect fluid: its
share leaves the anisotropic-stress budget while remaining in the total
radiation density, so that the numerator of Eq.~\eqref{eq:Rnu} loses
\rhoDG\ and $R_\nu=0.363$ instead. The two prescriptions therefore
bracket the treatment of the plasma, and the spread between them is a
systematic on the amplitude of the superhorizon initial conditions
rather than an adjustable quantity. We have not quantified its
propagation into the observables and do not rely on it being negligible;
the results reported below are obtained throughout with the
free-streaming prescription of Eq.~\eqref{eq:slaved}.
At \ac\ the BDE share is handed to the scalar field. As in
Ref.~\cite{Almaraz2019}, its perturbations obey
\begin{align}
  \delta\rho_{\rm BDE} &= \frac{\bar\phi'\delta\phi'}{a^2}+V_\phi\,\delta\phi ,
  \qquad
  \delta P_{\rm BDE} = \frac{\bar\phi'\delta\phi'}{a^2}-V_\phi\,\delta\phi ,
  \nonumber\\
  \delta\phi''&+2\mathcal{H}\,\delta\phi'+\left(k^2+a^2V_{\phi\phi}\right)\delta\phi
  =-\tfrac12\,\bar\phi'\,h' ,
  \label{eq:dphi}
\end{align}
with $V_\phi\equiv dV/d\phi$, and are matched to the plasma at \ac\
through $\delta_{\rm BDE}(\ac)=\delta_{\rm DG}(\ac)$ with
$\delta\phi'(\ac)=0$. With Eqs.~\eqref{eq:Vphi} and~\eqref{eq:rhobdeic}
this gives $\delta\phi(\ac)=-\tfrac92\,\fbde\,\Lc\,\delta_{\rm DG}(\ac)$,
the prescription of Ref.~\cite{Almaraz2019} for $\fbde=1$; for
inverse-power-law potentials the late-time solution is insensitive to
this initial value~\cite{Almaraz2019}.

\textit{The bound-dark-matter fluid.}---\label{sec:bdmfluid}%
After the transition, the BDM is a collisionless species with a narrow
momentum distribution centered on Eq.~\eqref{eq:va}. In principle its
perturbations should be evolved with a full Boltzmann hierarchy, as for
massive neutrinos. We instead use the imperfect-fluid description of the
earlier bound-dark-matter analysis~\cite{Mastache2020}, written in the
generalized-dark-matter form~\cite{Hu1998}, for two reasons: the momentum distribution is narrow and centered on a single
velocity rather than thermal and broad, so truncating at the shear is a
better approximation here than for neutrinos; and the component is already
semi-relativistic at condensation for the velocities of interest
($w_{\rm BDM}(\ac)=\vc^2/3\le1/6$, the bound saturating at the prior edge
$\vc=1/\sqrt2$) and becomes non-relativistic within a few
$e$-folds, so the regime in which higher multipoles matter is short.
The fluid variables are the density contrast $\delta$,
the velocity divergence $\theta$ and the shear $\sigma$:
\begin{align}
  \delta' &= -(1+w)\left(\theta+\frac{h'}{2}\right)
                -3\mathcal{H}\left(\ceff-w\right)\delta ,
  \label{eq:bdmcontinuity}\\[2pt]
  \theta' &= -\mathcal{H}\left(1-3\ca\right)\theta
                +\frac{\ceff}{1+w}\,k^2\delta - k^2\sigma ,
  \label{eq:bdmeuler}\\[2pt]
  \sigma' &= -3\left[\frac{1}{\tau}
                +\mathcal{H}\left(\frac{2}{3}-\ca-w\right)\right]\sigma
  \nonumber\\
             &\quad
                +\frac{8}{3}\,\frac{\cvis}{1+w}
                 \left(\theta+\frac{h'}{2}\right) ,
  \label{eq:bdmshear}
\end{align}
where $w\equiv w_{\rm BDM}$; the system is closed by
Eq.~\eqref{eq:ca2} together with
\begin{equation}
  \ceff=\ca,
  \qquad
  \cvis = 3\,w\,\ca ,
  \label{eq:closure}
\end{equation}
where the squared effective sound speed \ceff\ and the viscosity
parameter \cvis\ set the pressure support and the free-streaming
damping. Before the transition, where $w=1/3$ and $\ca=\cvis=1/3$,
Eqs.~\eqref{eq:bdmcontinuity}--\eqref{eq:bdmshear} reduce exactly to
Eqs.~(A1)--(A3) of Ref.~\cite{Mastache2020}, those of a massless,
neutrino-like component. After it, the Euler equation coincides with
their Eq.~(A5), and the continuity equation follows from the same
adiabatic sound speed, Eq.~\eqref{eq:ca2}. For the shear we take the
viscosity of the fluid approximation for non-cold
relics~\cite{LesgourguesTram2011}, $\cvis=3w\ca$, which coincides with
that of Ref.~\cite{Mastache2020} at the transition, and, as there, we
omit the $\eta'$ term of the shear source, $\eta$ being the second
synchronous-gauge potential~\cite{MaBertschinger1995}.

The BDM perturbations are integrated continuously through the
transition, with $w_{\rm BDM}(a)$ taken from Eq.~\eqref{eq:wbdm} and
smoothed over the matching window, so that
$\delta_{\rm BDM}(\ac)=\delta_{\rm DG}(\ac)$.
This is the appropriate prescription if the conversion is controlled by
the scale factor. Were it instead triggered by the local plasma density,
the transition hypersurface would be one of constant \rhoDG\ rather than
constant $a$, and the curvature perturbation inherited by each daughter
species would impose
$\delta_{\rm BDM}(\ac)=\tfrac34\big[1+w_{\rm BDM}(\ac)\big]\,\delta_{\rm DG}(\ac)$
instead, a $25\%$ reduction in the cold limit. The two prescriptions
coincide when $w_{\rm BDM}(\ac)=1/3$ and therefore never had to be distinguished
in the standalone analyses. We adopt the former throughout and record
the difference as a systematic on the small-scale amplitude; settling it
requires the small-scale analysis deferred to
Sec.~\ref{sec:discussion}.

\textit{Small-scale phenomenology.}---\label{sec:smallscales}%
The BDM imprints a suppression of the matter power spectrum involving
two nearly coincident scales. The first is the comoving horizon at the
transition, $k_{\rm hor}(\ac)=1/\tau(\ac)$, the mode $k_c$ of
Ref.~\cite{Almaraz2019}, of order
$1\,h\,\mathrm{Mpc}^{-1}$ for the preferred parameters: modes inside
the horizon before condensation see their dark-matter progenitor as
pressure-supported plasma and cannot grow. The second is the
free-streaming scale accumulated after the transition,
\begin{equation}
  \lambda_{\rm fs}
  = \int_{\ac}^{a_{\rm eq}} \frac{v(a)}{a^2 H(a)}\,da ,
  \label{eq:lambdafs}
\end{equation}
with $v(a)$ from Eq.~\eqref{eq:va} and $a_{\rm eq}$ the scale factor at
matter--radiation equality. The integral starts at \ac\ because the
constituents are massless before condensation, the scale they cover
while still part of the plasma being the horizon $1/k_{\rm hor}(\ac)$;
after \ac\ the velocity dispersion erases structure below
$\lambda_{\rm fs}$ in the manner familiar from warm dark matter.
Reference~\cite{Mastache2020} includes the relativistic phase in its
free-streaming length, which therefore exceeds ours by $1/k_{\rm hor}(\ac)$
at fixed \vc; we verified numerically that
the resulting cutoff scales with \vc\ as the analytic integral predicts.
The cutoff is summarized by the half-suppression scale \khalf,
\begin{equation}
  P_{\rm BDE\text{-}BDM}(\khalf)
  = \tfrac12\,P_{\vc\to0}(\khalf) ,
  \label{eq:khalf}
\end{equation}
whose fitted value is quoted in Sec.~\ref{sec:pkresult}.

The abruptness of the transition distinguishes this component from
thermal warm dark matter. In a thermal scenario, the mass is constant and
the velocity redshifts smoothly, so the passage to the non-relativistic
regime is gradual; in BDM the mass is acquired non-perturbatively at
\ac\ and the velocity drops abruptly to \vc, so the free-streaming
length accumulated is smaller for the same epoch of non-relativistic
transition. Reference~\cite{Mastache2020} quantifies this: a $3$~keV
thermal particle gives
$M_{\rm fs}=1.97\times10^{7}\,M_\odot\,h^{-3}$, whereas a BDM particle
with the same scale factor $a_{\rm nr}$ at which it becomes
non-relativistic, but an abrupt transition at $\vc\simeq0$
gives $2.08\times10^{4}\,M_\odot\,h^{-3}$, three orders of magnitude
smaller. The velocity therefore interpolates between two qualitatively
different regimes, and the model cannot be mapped onto an effective
warm-dark-matter mass without specifying it. Because $k_{\rm hor}(\ac)$
and the free-streaming cutoff nearly coincide, modes just beyond the
cutoff are doubly suppressed and the spectrum falls more steeply than a
thermal cutoff with the same \khalf. Deep in the suppressed tail the
transfer function retains oscillations from the acoustic phase at
erasure; these lie where the power is suppressed by more than an order
of magnitude and where the fluid description is no longer controlled, so
we exclude $k>1.5\,h\,\mathrm{Mpc}^{-1}$ from all analyses and figures.

On larger scales the model produces a smooth offset from \LCDM\ whose
sign comes from a competition between the two dark-sector components.
The bound-dark-energy sector alone \emph{enhances} small-scale power: a
mode entering the horizon before \ac\ first loses amplitude to the free
streaming of the still-massless dark-group particles, then regains more
than it lost once the condensate dilutes as
$\rho_{\rm BDE}\propto a^{-6}$. In BDE-CDM, where the dark matter is
cold and this is the only dark-sector effect, the outcome is a
progressive enhancement reaching approximately $25\%$ near
$k\approx4.3\,\mathrm{Mpc}^{-1}$ ($\simeq6.4\,h\,\mathrm{Mpc}^{-1}$)~\cite{MacorraLozano2025},
the scale at which the earlier constraints of Ref.~\cite{Almaraz2019}
gave $18\%$. In the
unified model, the bound dark matter acts oppositely on those same
scales, and for the preferred velocity its free-streaming suppression
exceeds the BDE enhancement, so the net deviation is negative and grows
with $k$ through the cutoff. Integrated over the standard window this
lowers the amplitude of structure, which is the origin of the reduced
\sig, the rms linear density fluctuation in spheres of radius
$8\,h^{-1}\,\mathrm{Mpc}$, and $\Seight\equiv\sig(\Om/0.3)^{1/2}$ reported
in Sec.~\ref{sec:vcresult}.

This feature is the reason for introducing the bound dark matter, since
removing power on sub-galactic scales thins the population of low-mass
halos. That was the original purpose of the
component~\cite{Macorra2010,Mastache2020}, and it is a motivation rather
than a result of the present analysis: the scales relevant to satellite
abundances, $k\sim10$--$50\,h\,\mathrm{Mpc}^{-1}$, lie far beyond the
range on which our likelihoods draw, and we make no claim here about
satellite counts. What this analysis does establish is that the unified
model inherits two opposite small-scale behaviors from its two sectors
and that \vc\ controls which dominates: in the cold limit the BDE
enhancement is recovered, while for the velocities our data do not
exclude the free-streaming suppression wins. Measurements resolving
$k\gtrsim1\,h\,\mathrm{Mpc}^{-1}$ therefore discriminate not only
between BDE-BDM and \LCDM\ but between the unified model and its own
cold limit.

\section{Data and methodology}
\label{sec:data}

All models are constrained with the same measurements, described
as follows:

\begin{itemize}
\item \textit{DESI}: BAO distances and expansion rates from galaxies,
quasars and Lyman-$\alpha$ tracers, from the second data release
(DR2)~\cite{DESI2025II,DESI2025BAO}.

\item \textit{Planck}: temperature and polarization spectra at
$\ell>30$ from the NPIPE CamSpec likelihood~\cite{Planck2020VI,%
Rosenberg2022}, and the low-$\ell$ TT and EE likelihoods at
$2\le\ell\le30$~\cite{Planck2020VI,Planck2020V}.

\item \textit{CMB lensing}: the combination of the Planck PR4 and ACT DR6
lensing reconstructions~\cite{ACTDR6map}, through the ACT DR6 lensing
likelihood~\cite{ACTDR6lens}.

\item \textit{PantheonPlus}: 1701 light curves of 1550
spectroscopically confirmed SNe Ia over $0<z<2.3$~\cite{Scolnic2022}.

\item \textit{Union3.1}: SNe Ia compiled from a range of surveys and
standardized within the UNITY Bayesian framework~\cite{Rubin2023}.

\item \textit{DES-Dovekie}: the DES five-year sample with a revised
low-redshift calibration, which yields a weaker preference for evolving
dark energy than the original release~\cite{DESDovekie}.
\end{itemize}

We label the combination of the first three BAO $+$ CMB throughout,
leaving the lensing term implicit, and add one supernova compilation at
a time, reporting the three separately because the strength of the
preference for evolving dark energy depends on the
sample~\cite{DESI2025II}. In the text we quote the Union3.1 combination,
with PantheonPlus as a complement; the companion
letter~\cite{LozanoMacorra2026PRL} quotes the DES-Dovekie combination,
and Tables~\ref{tab:full}, \ref{tab:dic_full} and~\ref{tab:chi2_all}
report all three.

Observables are computed with a modified version of
\textsc{camb}~\cite{Lewis2000} driven by
\textsc{cobaya}~\cite{Torrado2021}. The dark-group background, the
scalar field with the initial conditions of Eqs.~\eqref{eq:rhobdeic}
and~\eqref{eq:phidotc}, the partition of Eq.~\eqref{eq:fbde}, and the
bound-dark-matter fluid of Sec.~\ref{sec:bdmfluid} are implemented
natively, and the transient radiation of Eq.~\eqref{eq:Next} enters the
expansion rate, the adiabatic initial conditions and the BBN helium
prediction consistently.

Nothing in the dark-energy sector plays the role of $\rho_\Lambda$ in
\LCDM: the present dark-energy density is the endpoint of integrating
Eq.~\eqref{eq:kg} from the initial data
\eqref{eq:rhobdeic}--\eqref{eq:phidotc}, and $\Hub$ likewise follows
from the background solution, so sampling $\Hub$ or $\theta_{\rm MC}$ (the \textsc{camb}
approximation to the angular size of the sound horizon at recombination)
alongside \Lc\ would over-specify the system. We therefore sample \Lc\
in place of $\theta_{\rm MC}$, as in the companion BDE-CDM
analysis~\cite{MacorraLozano2025}, use $\omega_{\rm bdm}$ in place of
$\omega_c$, and add the new quantity, \vc. The remaining
dark-sector quantities are derived: \ac\ from Eq.~\eqref{eq:acLc} and
\fbde\ from Eq.~\eqref{eq:fbde}. The sampled parameters are therefore
\begin{equation}
  \mathcal{P}_{\rm BDE\text{-}BDM}
  = \bigl\{\omega_b,\,\omega_{\rm bdm},\,\tau_{\rm reio},\,n_s,\,
           \ln(10^{10}A_s),\,\Lc,\,\vc\bigr\},
  \label{eq:params}
\end{equation}
seven parameters, $\tau_{\rm reio}$ being the optical depth to
reionization: one more than flat \LCDM, because of \vc, and one
fewer than \wowa.

The priors are uniform and are listed in Table~\ref{tab:priors}. Two are
specific to the unified model. On \Lc\ we impose
$\mathcal{U}[23,\,50]$~eV, chosen to span the theoretical prediction
$\Lc^{\rm th}=34^{+16}_{-11}$~eV of Eq.~\eqref{eq:Lc}. On the other hand, on \vc\ we impose
$\mathcal{U}[10^{-3},\,1/\sqrt2]$, the upper limit being the velocity at
which the momentum of the bound states equals their mass and the lower a
numerical floor. The remaining priors follow the DESI DR2
analysis~\cite{DESI2025II}, and the neutrino mass is fixed to
$\sum m_\nu=0.06$~eV. The same data, likelihood versions, and
settings are used for \LCDM\ and \wowa.

Sampling continued until the Gelman--Rubin statistic satisfied
$R-1<0.01$. Best-fit values were obtained with the BOBYQA
minimizer~\cite{Powell2009}; for \LBDM\ we use the lowest deviance found
in the chains.

\begin{table}[t]
\caption{\label{tab:priors}%
Uniform priors on the sampled cosmological parameters. Entries marked
``--'' are not sampled in the corresponding model. The derived
parameters \ac, \fbde, and $\Hub$ follow from Eqs.~\eqref{eq:acLc} and
\eqref{eq:fbde} and from the background evolution, respectively, and
carry no priors of their own. BDE-CDM uses the BDE-BDM priors with
$\omega_c$ in place of $\omega_{\rm bdm}$ and without \vc; \LBDM\ and
\wowaBDM\ use the \LCDM\ and \wowa\ priors with $\omega_{\rm bdm}$ in place
of $\omega_c$, $\log_{10}\ac\in\mathcal{U}[-12,\,-4]$ and
$\vc\in\mathcal{U}[10^{-3},\,0.99]$.}
\begin{ruledtabular}
\begin{tabular}{l l l}
Parameter & \LCDM, \wowa & BDE-BDM \\
\colrule
$\ln(10^{10}A_s)$      & $\mathcal{U}[1.61,\,3.91]$ & $\mathcal{U}[1.61,\,3.91]$ \\
$n_s$                  & $\mathcal{U}[0.8,\,1.2]$   & $\mathcal{U}[0.8,\,1.2]$ \\
$100\,\theta_{\rm MC}$ & $\mathcal{U}[0.5,\,10.0]$  & -- \\
$\Lc$~[eV]             & --                         & $\mathcal{U}[23,\,50]$ \\
$\omega_b$             & $\mathcal{U}[0.005,\,0.1]$ & $\mathcal{U}[0.005,\,0.1]$ \\
$\omega_c$             & $\mathcal{U}[0.001,\,0.99]$ & -- \\
$\omega_{\rm bdm}$     & --                         & $\mathcal{U}[0.001,\,0.99]$ \\
$\tau_{\rm reio}$       & $\mathcal{U}[0.01,\,0.8]$  & $\mathcal{U}[0.01,\,0.8]$ \\
$\vc$                  & --                         & $\mathcal{U}[10^{-3},\,1/\sqrt2]$ \\
$w_0$                  & $\mathcal{U}[-3,\,1]$      & -- \\
$w_a$                  & $\mathcal{U}[-3,\,2]$      & -- \\
\end{tabular}
\end{ruledtabular}
\end{table}

\section{Results}
\label{sec:results}
\begin{figure*}[!htbp]
\includegraphics[width=0.86\textwidth]{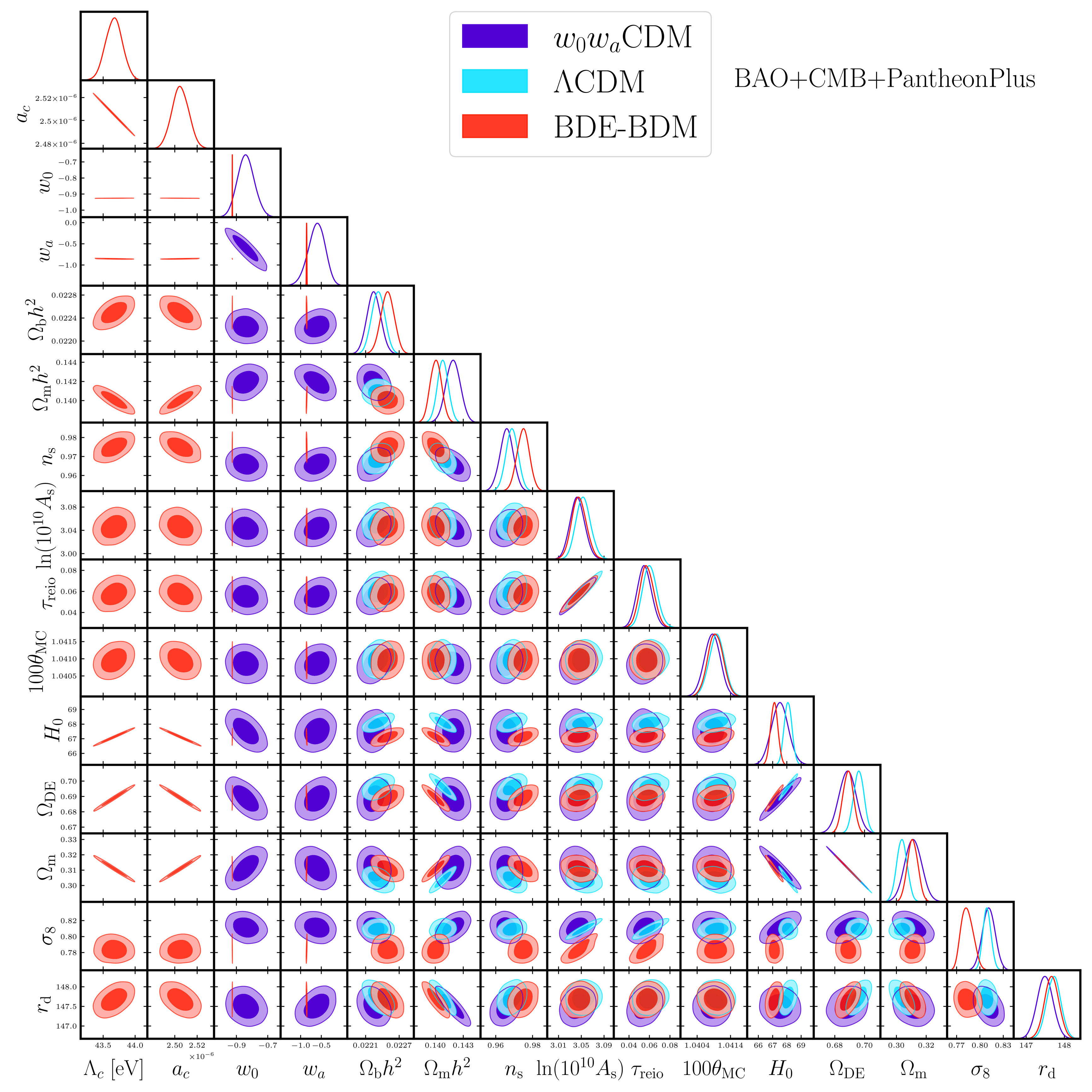}
\caption{\label{fig:triangle}%
Marginalized posteriors and $68\%$/$95\%$ confidence contours for
BDE-BDM, \LCDM\ and \wowa\ from DESI DR2 BAO $+$ CMB $+$ PantheonPlus
$+$ CMB lensing, in the plane of
$(\Lc,\ac,w_0,w_a,\omega_b,\omega_m,n_s,\ln(10^{10}A_s),
\tau_{\rm reio},100\theta_{\rm MC},\Hub,\Ode,\Om,\sig,r_{\rm d})$. The $w_0$ and
$w_a$ posteriors of BDE-BDM appear as vertical lines on this scale,
being narrower than the \wowa\ ones by some four orders of magnitude. The
corresponding plots for Union3.1 and DES-Dovekie are quantitatively
similar and are summarized in Table~\ref{tab:full}.}
\end{figure*}

\begin{table*}[!tbp]
\caption{\label{tab:full}%
Marginalized constraints means and $68\%$ credible intervals on the baseline and derived cosmological
parameters of the BDE-BDM, \wowa, and \LCDM\ models. Each block corresponds to DESI DR2 BAO and CMB data combined
with one of the three SNe~Ia compilations.}
\begin{ruledtabular}
\scriptsize
\setlength{\tabcolsep}{4pt}
\begin{tabular}{l c c c}
Parameter & BDE-BDM & \wowa & \LCDM \\
\colrule
\multicolumn{4}{c}{DESI DR2 BAO $+$ CMB $+$ PantheonPlus}\\
\colrule
$\Lc$~[eV] & $43.67 \pm 0.13$ & -- & -- \\
$10^{6}\,\ac$ & $2.5050 \pm 0.0076$ & -- & -- \\
$\vc$ & $0.474 \pm 0.168$ & -- & -- \\
$100\,\theta_{\rm MC}$ & $1.04095 \pm 0.00024$ & $1.04085 \pm 0.00024$ & $1.04099 \pm 0.00023$ \\
$\omega_b$ & $0.02249 \pm 0.00012$ & $0.02225 \pm 0.00012$ & $0.02233 \pm 0.00012$ \\
$\omega_c$ & -- & $0.11901 \pm 0.00081$ & $0.11786 \pm 0.00061$ \\
$\omega_{\rm bdm}$ & $0.11693 \pm 0.00063$ & -- & -- \\
$\ln(10^{10}A_s)$ & $3.046 \pm 0.013$ & $3.043 \pm 0.013$ & $3.054 \pm 0.013$ \\
$n_s$ & $0.9749 \pm 0.0034$ & $0.9659 \pm 0.0036$ & $0.9689 \pm 0.0033$ \\
$\tau_{\rm reio}$ & $0.0575 \pm 0.0071$ & $0.0551 \pm 0.0072$ & $0.0605 \pm 0.0073$ \\
\colrule
$w_0$ & $-0.92611 \pm 0.00048$ & $-0.838 \pm 0.055$ & $-1$ \\
$w_a$ & $-0.8539 \pm 0.0054$ & $-0.61 \pm 0.21$ & $0$ \\
$\Hub$~[km\,s$^{-1}$\,Mpc$^{-1}$] & $67.12 \pm 0.26$ & $67.52 \pm 0.59$ & $68.08 \pm 0.28$ \\
$\Om$ & $0.3109 \pm 0.0035$ & $0.3114 \pm 0.0057$ & $0.3038 \pm 0.0035$ \\
$\Ode$ & $0.6890 \pm 0.0035$ & $0.6886 \pm 0.0057$ & $0.6961 \pm 0.0035$ \\
$\sig$ & $0.7836 \pm 0.0077$ & $0.8113 \pm 0.0082$ & $0.8092 \pm 0.0054$ \\
$\Seight$ & $0.7977 \pm 0.0095$ & $0.8265 \pm 0.0083$ & $0.8144 \pm 0.0074$ \\
$r_{\rm d}$~[Mpc] & $147.67 \pm 0.19$ & $147.50 \pm 0.21$ & $147.72 \pm 0.18$ \\
$\fbde$ & $0.9353 \pm 0.0062$ & -- & -- \\
\colrule
\multicolumn{4}{c}{DESI DR2 BAO $+$ CMB $+$ Union3.1}\\
\colrule
$\Lc$~[eV] & $43.67 \pm 0.13$ & -- & -- \\
$10^{6}\,\ac$ & $2.5052 \pm 0.0077$ & -- & -- \\
$\vc$ & $0.463 \pm 0.171$ & -- & -- \\
$100\,\theta_{\rm MC}$ & $1.04096 \pm 0.00023$ & $1.04082 \pm 0.00024$ & $1.04100 \pm 0.00023$ \\
$\omega_b$ & $0.02249 \pm 0.00012$ & $0.02223 \pm 0.00012$ & $0.02233 \pm 0.00012$ \\
$\omega_c$ & -- & $0.11928 \pm 0.00084$ & $0.11778 \pm 0.00061$ \\
$\omega_{\rm bdm}$ & $0.11696 \pm 0.00064$ & -- & -- \\
$\ln(10^{10}A_s)$ & $3.045 \pm 0.013$ & $3.041 \pm 0.013$ & $3.054 \pm 0.013$ \\
$n_s$ & $0.9746 \pm 0.0034$ & $0.9652 \pm 0.0036$ & $0.9690 \pm 0.0033$ \\
$\tau_{\rm reio}$ & $0.0570 \pm 0.0069$ & $0.0540 \pm 0.0071$ & $0.0608 \pm 0.0072$ \\
\colrule
$w_0$ & $-0.92615 \pm 0.00048$ & $-0.736 \pm 0.082$ & $-1$ \\
$w_a$ & $-0.8535 \pm 0.0053$ & $-0.92 \pm 0.28$ & $0$ \\
$\Hub$~[km\,s$^{-1}$\,Mpc$^{-1}$] & $67.11 \pm 0.26$ & $66.65 \pm 0.79$ & $68.12 \pm 0.28$ \\
$\Om$ & $0.3111 \pm 0.0036$ & $0.3201 \pm 0.0079$ & $0.3036 \pm 0.0035$ \\
$\Ode$ & $0.6889 \pm 0.0036$ & $0.6798 \pm 0.0079$ & $0.6966 \pm 0.0035$ \\
$\sig$ & $0.7817 \pm 0.0083$ & $0.8055 \pm 0.0091$ & $0.8092 \pm 0.0053$ \\
$\Seight$ & $0.796 \pm 0.010$ & $0.8319 \pm 0.0089$ & $0.8137 \pm 0.0074$ \\
$r_{\rm d}$~[Mpc] & $147.67 \pm 0.19$ & $147.45 \pm 0.21$ & $147.73 \pm 0.19$ \\
$\fbde$ & $0.9357 \pm 0.0061$ & -- & -- \\
\colrule
\multicolumn{4}{c}{DESI DR2 BAO $+$ CMB $+$ DES-Dovekie}\\
\colrule
$\Lc$~[eV] & $43.68 \pm 0.13$ & -- & -- \\
$10^{6}\,\ac$ & $2.5045 \pm 0.0076$ & -- & -- \\
$\vc$ & $0.470 \pm 0.164$ & -- & -- \\
$100\,\theta_{\rm MC}$ & $1.04097 \pm 0.00023$ & $1.04083 \pm 0.00024$ & $1.04098 \pm 0.00023$ \\
$\omega_b$ & $0.02250 \pm 0.00012$ & $0.02224 \pm 0.00013$ & $0.02232 \pm 0.00012$ \\
$\omega_c$ & -- & $0.11917 \pm 0.00081$ & $0.11791 \pm 0.00060$ \\
$\omega_{\rm bdm}$ & $0.11690 \pm 0.00062$ & -- & -- \\
$\ln(10^{10}A_s)$ & $3.046 \pm 0.013$ & $3.042 \pm 0.013$ & $3.054 \pm 0.013$ \\
$n_s$ & $0.9750 \pm 0.0033$ & $0.9654 \pm 0.0036$ & $0.9687 \pm 0.0033$ \\
$\tau_{\rm reio}$ & $0.0574 \pm 0.0071$ & $0.0545 \pm 0.0070$ & $0.0603 \pm 0.0072$ \\
\colrule
$w_0$ & $-0.92611 \pm 0.00046$ & $-0.806 \pm 0.057$ & $-1$ \\
$w_a$ & $-0.8539 \pm 0.0051$ & $-0.73 \pm 0.22$ & $0$ \\
$\Hub$~[km\,s$^{-1}$\,Mpc$^{-1}$] & $67.13 \pm 0.26$ & $67.32 \pm 0.55$ & $68.06 \pm 0.27$ \\
$\Om$ & $0.3107 \pm 0.0035$ & $0.3135 \pm 0.0054$ & $0.3042 \pm 0.0035$ \\
$\Ode$ & $0.6892 \pm 0.0035$ & $0.6864 \pm 0.0054$ & $0.6958 \pm 0.0035$ \\
$\sig$ & $0.7816 \pm 0.0084$ & $0.8106 \pm 0.0079$ & $0.8093 \pm 0.0053$ \\
$\Seight$ & $0.795 \pm 0.010$ & $0.8285 \pm 0.0083$ & $0.8149 \pm 0.0074$ \\
$r_{\rm d}$~[Mpc] & $147.68 \pm 0.19$ & $147.47 \pm 0.21$ & $147.71 \pm 0.18$ \\
$\fbde$ & $0.9356 \pm 0.0060$ & -- & -- \\
\end{tabular}
\end{ruledtabular}
\end{table*}
\begin{figure*}[!htbp]
\includegraphics[width=0.86\textwidth]{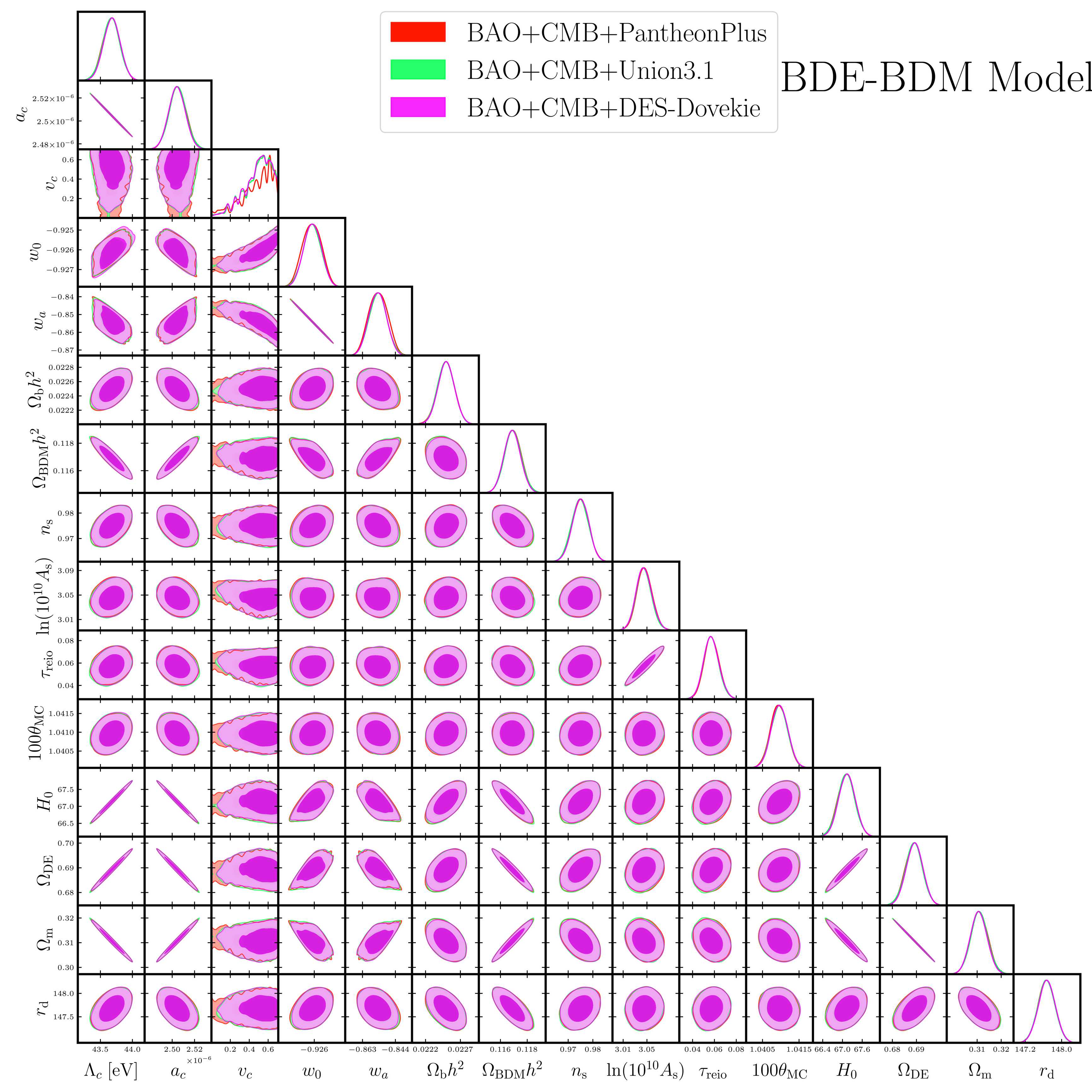}
\caption{\label{fig:triangle_sne}%
Marginalized posteriors of the BDE-BDM model, at $68\%$ and $95\%$, for
DESI DR2 BAO $+$ CMB $+$ CMB lensing combined with each of the three
supernova compilations. The tight anticorrelation between \Lc\ and \ac\
is the condensation relation Eq.~\eqref{eq:acLc}, not a data-driven
degeneracy.}
\end{figure*}
The marginalized $68\%$ confidence constraints on the sampled parameters
and on some selected derived quantities are listed in
Table~\ref{tab:full}, for the three cosmological models and the three dataset
combinations, and the corresponding posteriors and confidence contours
are shown in Fig.~\ref{fig:triangle}.

Throughout this section, we exhibit figures for two of the three supernova
compilations. Union3.1 gives the strongest preference for evolving dark
energy in the DESI DR2 analysis and PantheonPlus the weakest;
DES-Dovekie falls between them, and its figures add nothing that the two
other SNe compilations do not already show in the plots. However, numerical results for all three SNe compilations
are given in Tables~\ref{tab:full} and \ref{tab:dic_full}.

\subsection{The condensation scale}
\label{sec:lcresult}

Sampling \Lc\ within the prior of Sec.~\ref{sec:data}, we obtain for the condensation energy scale \Lc\ the value of
\begin{equation}
  \Lc = 43.67\pm0.13~\mathrm{eV}
  \qquad(\text{BAO}+\text{CMB}+\text{Union3.1}),
  \label{eq:Lcfit}
\end{equation}
together with the implied transition epoch, which follows from
Eq.~\eqref{eq:acLc},
\begin{equation}
  \ac = (2.505\pm0.008)\times10^{-6},
  \label{eq:acfit}
\end{equation}
and the condensing fraction $\fbde=0.936\pm0.006$ from
Eq.~\eqref{eq:fbde}. The dark group therefore delivers
$\Omega_{\rm BDE}(\ac)=0.1052$ to the meson and
$\Omega_{\rm BDM}(\ac)=0.0072$ to the massive states, out of a total
$\Omega_{\rm DG}(\ac)=0.1125$, and Eq.~\eqref{eq:phidotc} places the
initial equation of state of the field at
$w_{\rm BDE}(\ac)=0.287$, in agreement with the value recovered
from the chains. The energy density carried into the massive states,
Eq.~\eqref{eq:rhobdmc}, corresponds to a transition energy scale
$\rho_{\rm BDM}(\ac)^{1/4}=28.9$~eV, and \fbde\ sits comfortably above
the lower endpoint of the physical bound Eq.~\eqref{eq:fbdebound}, which
would be reached only for $\omega_{\rm bdm}$ an order of magnitude
larger than observed. The recovered value of \Lc\ lies in the interior of
the prior range $\mathcal{U}[23,50]$~eV set by Eq.~\eqref{eq:Lc}, at
$0.6\sigma$ above the central value of the theoretical prediction in
units of the theoretical uncertainty, and is stable across the three SNe
compilations to $0.012$~eV, an order of magnitude inside the
statistical error (Table~\ref{tab:full}). Exchanging one supernova sample for
another therefore does not move the inferred \Lc.

The BDE-CDM model~\cite{MacorraLozano2025}, fitted to the same data,
gives $\Lc=43.76\pm0.13$~eV and $\ac=(2.500\pm0.008)\times10^{-6}$,
$0.09$~eV above the value found here. The two need not coincide, and
the direction of the shift is expected. In BDE-CDM the entire dark group
condenses into the meson, $\fbde=1$, and the cold dark matter is a
separate component present from BDE. In the unified model, a
fraction $1-\fbde=0.064$ of the dark-group energy is passed into the
bound dark matter at the transition, so the scalar field starts with
$6.4\%$ less energy and with the correspondingly modified initial
condition Eq.~\eqref{eq:phidotc}; reproducing the same present
dark-energy density therefore requires a slightly smaller condensation
scale. The dark matter of the unified model is moreover part of the dark
group before \ac, and thus contributes to the radiation budget rather
than to the matter budget at early times, which further distinguishes it
from BDE-CDM even in the cold limit $\vc\to0$
(Sec.~\ref{sec:coldlimit}).

Equation~\eqref{eq:acfit} follows from \Lc\
through the gauge relation Eq.~\eqref{eq:acLc}, whereas in the
standalone bound-dark-matter analyses the transition scale is an
independent quantity. Reference~\cite{Macorra2010} required a
transition energy scale of $\mathcal{O}(10\text{--}100)$~eV on
structure-formation grounds alone; the value implied here,
$\rho_{\rm BDM}(\ac)^{1/4}=28.9$~eV from Eq.~\eqref{eq:rhobdmc} with
Eq.~\eqref{eq:Lcfit} and $\fbde=0.936$, falls within that range without
having been adjusted to it, having been obtained instead from
gauge-coupling unification together with the energy partition at
condensation. Reference~\cite{Mastache2020}, scanning the pair
$(\ac,\vc)$ against CMB, SNe~Ia, and BAO data, placed the $1\sigma$ upper
bound $\ac<2.66\times10^{-6}$ for $\vc\to0$, which tightens with the
velocity, to $\ac<4.59\times10^{-7}$ at $\vc=1/\sqrt2$, because the
relativistic phase must not alter the radiation content at
matter--radiation equality. Eq.~\eqref{eq:acfit} satisfies the bound in
the cold limit, and it is placed there by the separate requirement that
the meson reproduce the observed dark-energy density.

Figure~\ref{fig:triangle_sne} shows the confidence contours across the full
parameter space. The posterior of the model is plotted for the three
supernova compilations overlaid, and the contours coincide almost
exactly in every direction except \vc: \Lc, \ac, $w_0$, $w_a$, $\Hub$,
\Om\ and the sound horizon at the drag epoch, $r_{\rm d}$, are recovered
identically whichever sample is used.
A model whose dark-energy sector is fixed by the gauge dynamics rather
than by the low-redshift distance data should behave in this way, and
the one parameter that is not so fixed, the bound velocity, is also the
one where the compilations visibly differ.

\begin{figure*}[!htbp]
\includegraphics[width=0.336\textwidth]{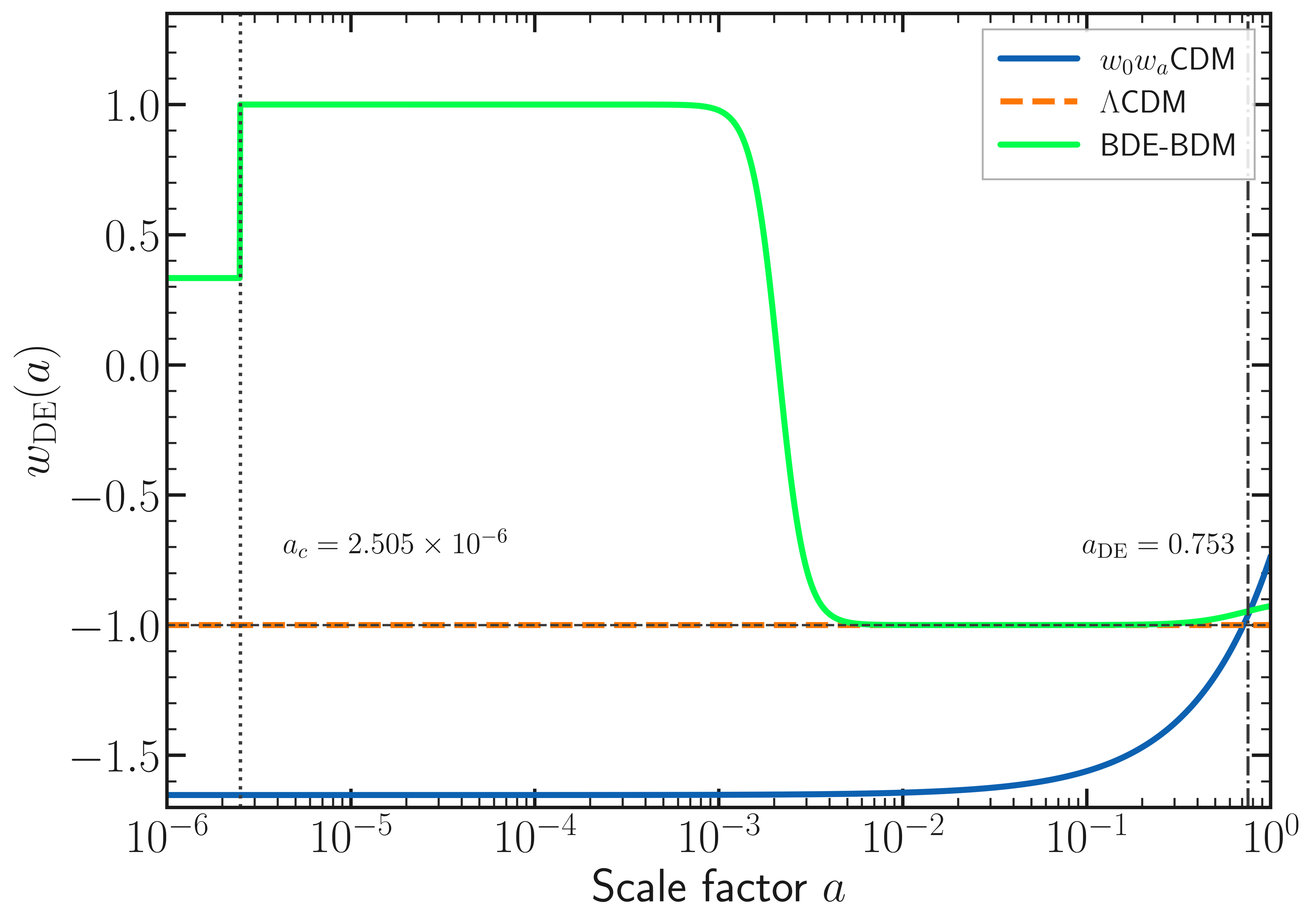}\hfill
\includegraphics[width=0.338\textwidth]{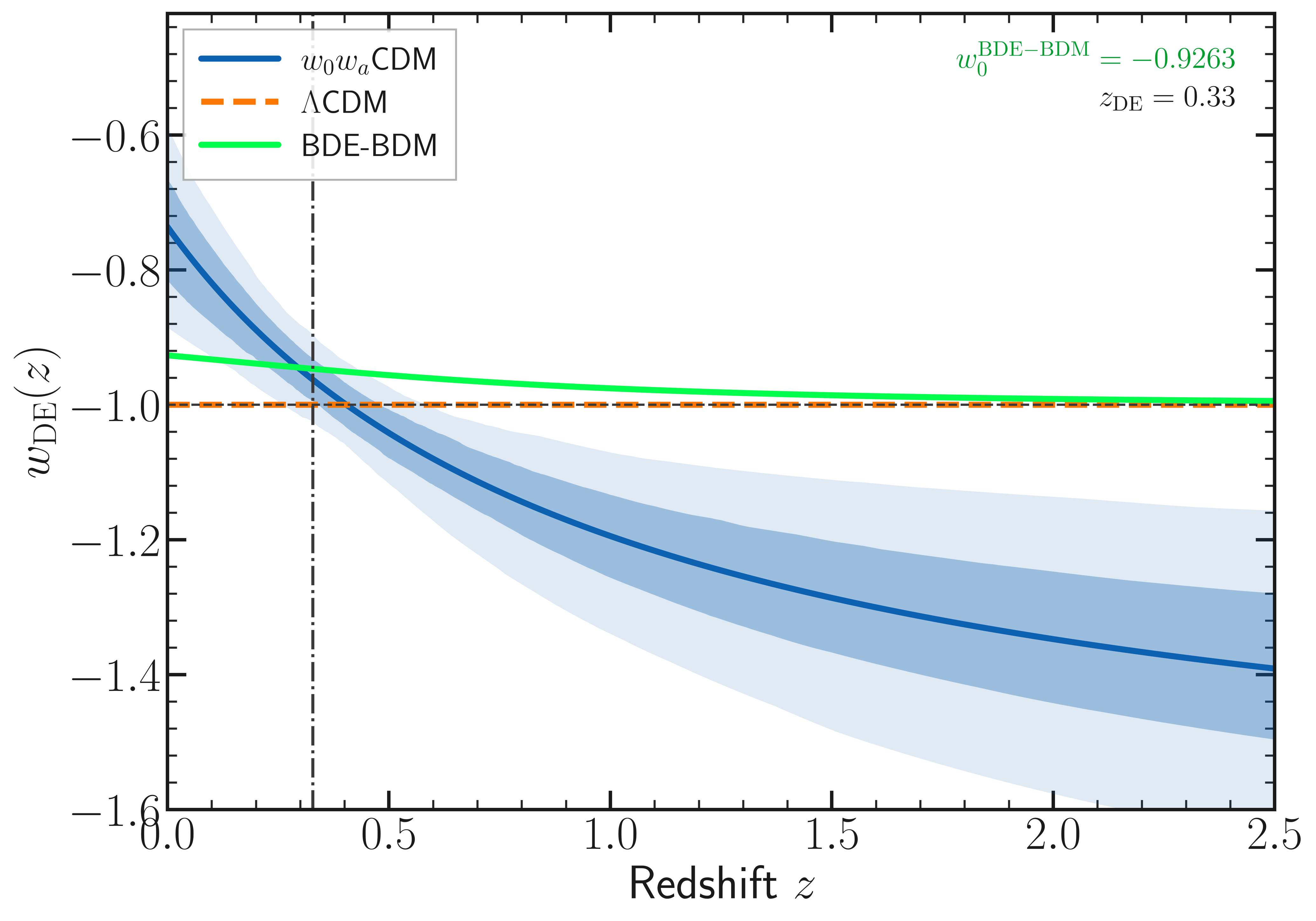}\hfill
\includegraphics[width=0.305\textwidth]{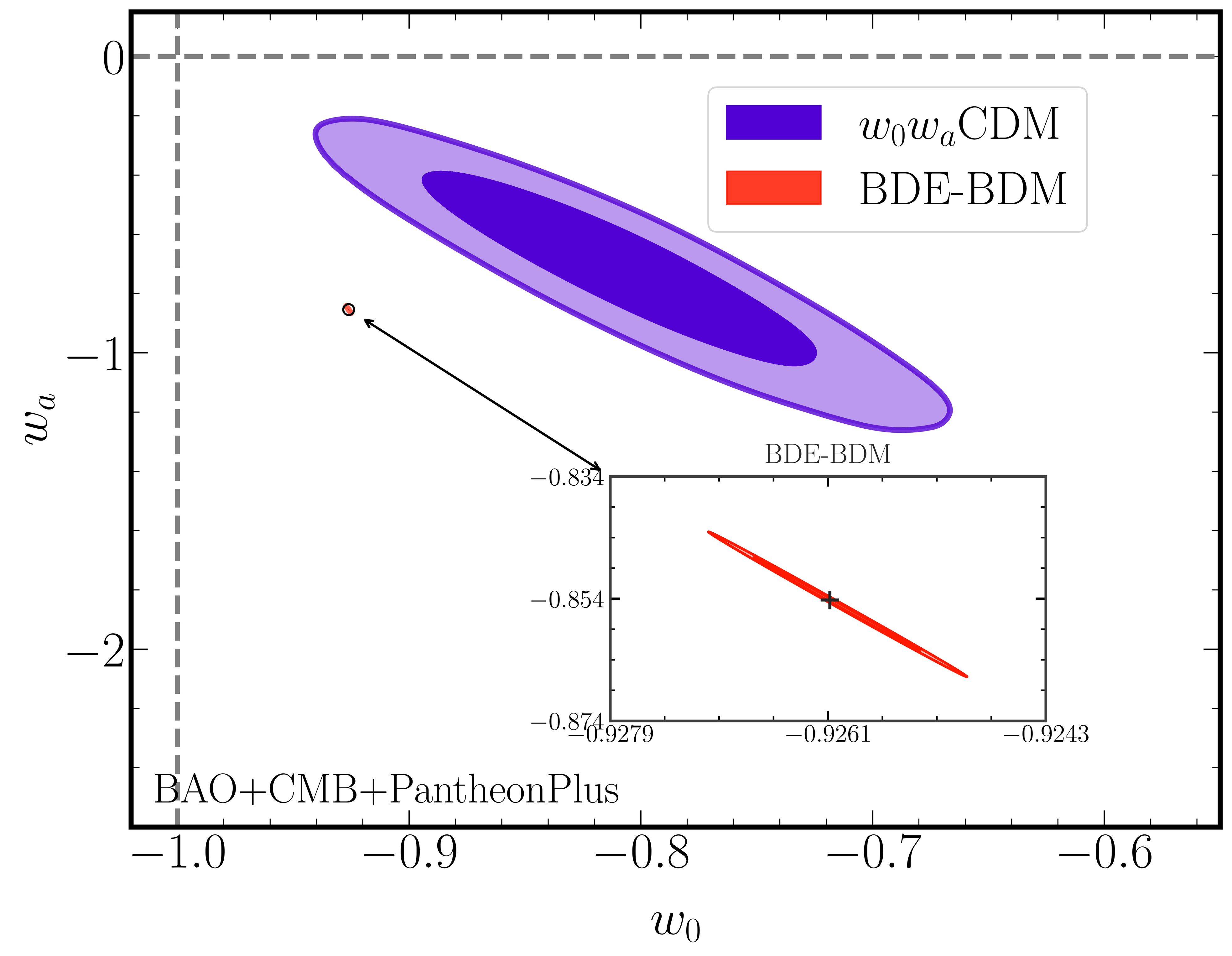}
\caption{\label{fig:background}%
Equation of state of the dark energy and its projection onto the $w_0w_a$
plane. \textit{Left} and \textit{center}: $w_{\rm DE}$ against scale
factor and against redshift, at the marginalized means of DESI DR2 BAO
$+$ CMB $+$ Union3.1 $+$ CMB lensing. Dotted and dash-dotted lines
mark condensation at $\ac=2.505\times10^{-6}$ and matter--dark-energy
equality at $a_{\rm DE}=0.753$ ($z_{\rm DE}=0.33$); the BDE-BDM band is narrower than the line width.
\textit{Right}: the $(w_0,w_a)$ plane for DESI DR2 BAO
$+$ CMB $+$ PantheonPlus $+$ CMB lensing, the compilation
giving the weakest preference for evolving dark energy. Filled contours
are the $68\%$ and $95\%$ \wowa\ constraints; the point is the BDE-BDM
prediction, magnified in the zoom plot for clarity.}
\end{figure*}

\subsection{Background constraints}
\label{sec:background}

For the baseline data plus Union3.1 SNe the unified model gives
\begin{align}
  \Hub    &= 67.11\pm0.26~\mathrm{km\,s^{-1}\,Mpc^{-1}}, \nonumber\\
  \Om     &= 0.311\pm0.004, \qquad
  \Ode     = 0.689\pm0.004, \nonumber\\
  \omega_{\rm bdm} &= 0.1170\pm0.0006 . \nonumber
\end{align}
Three features of Table~\ref{tab:full} are worth noting; the history of
the equation of state that underlies them is analyzed in
Sec.~\ref{sec:eos}.

First, the expansion-history parameters are close to those of \LCDM\ for
the same data ($\Hub=68.12\pm0.28$, $\Om=0.304\pm0.004$): the model
reproduces the measured distances with its predicted $w(z)$, without
requiring free parameters in the dark-energy sector.

Second, \Hub\ is determined three times more precisely than under \wowa{}
($\Hub=66.65\pm0.79$), which follows from the absence of freedom rather
than from any additional information: because \Lc\ already fixes the
dark-energy history, the degeneracy between $\Hub$ and $w$ that widens
the \wowa\ interval never opens. The
same rigidity is visible in the present equation of state,
Eq.~\eqref{eq:w0waproj} below, whose uncertainty is two orders of
magnitude smaller than the \wowa\ constraint $w_0=-0.736\pm0.082$ obtained
from the same data. Once again, the narrowness measures how tightly the theory binds
$w(z)$ to \Lc, not how sharply the data discriminate.

Third, the sound horizon $r_{\rm d}$ and the drag redshift $z_{\rm d}$ are
essentially unchanged between the three models ($r_{\rm d}=147.67\pm0.19$~Mpc for BDE-BDM
against $147.73\pm0.19$~Mpc for \LCDM), which confirms that the
pre-recombination physics is not modified in any way that the CMB can
detect and that all the differences reside in the late-time sector. The
small shift in $z_{\rm d}$ ($1060.39\pm0.27$ against
$1059.69\pm0.27$) reflects the transient dark radiation slightly
advancing the drag epoch.

\begin{figure}[!htbp]
\includegraphics[width=\columnwidth]{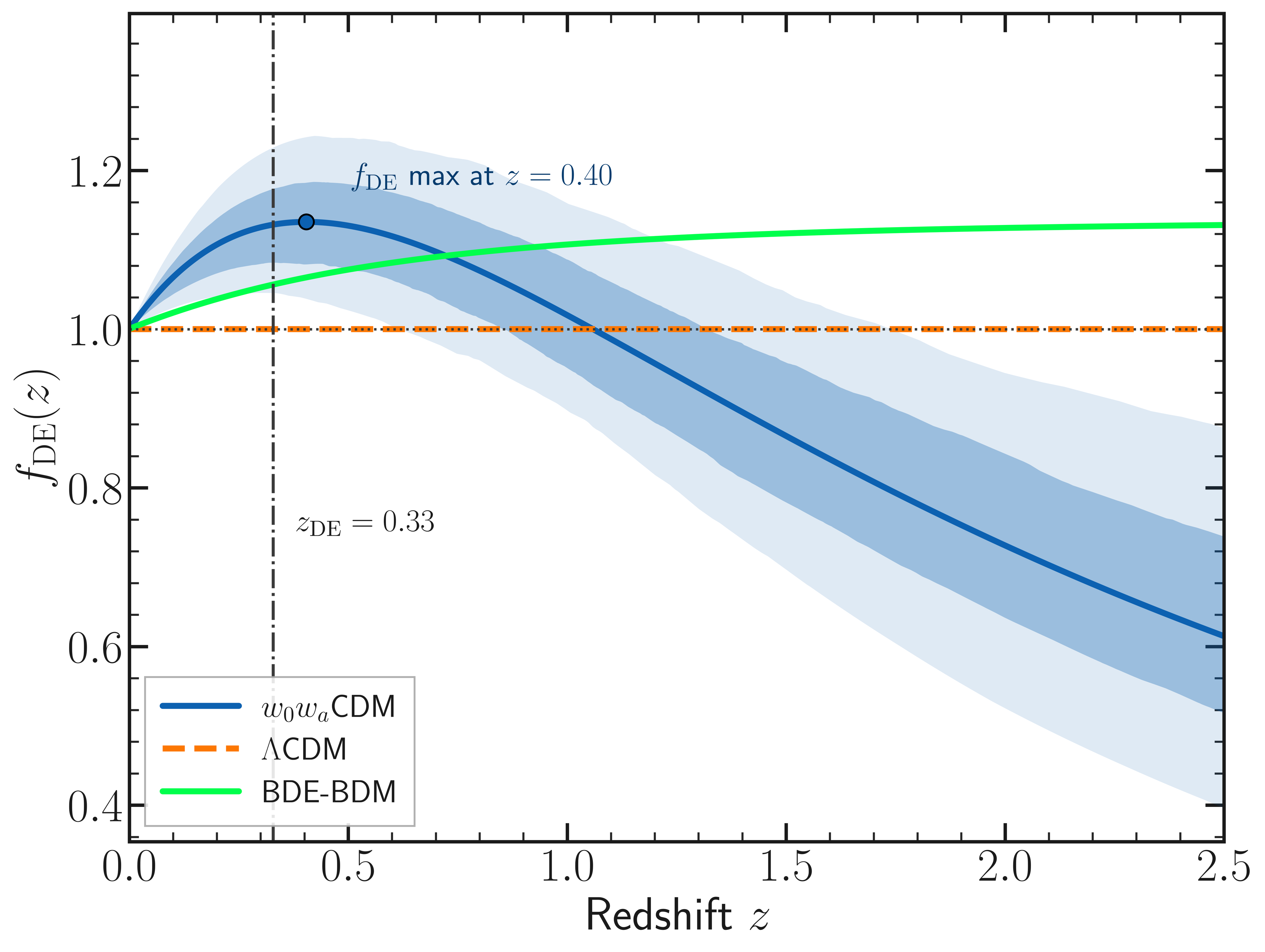}
\caption{\label{fig:fde}%
Dark-energy density normalized to its present value,
$f_{\rm DE}(z)=\rho_{\rm DE}(z)/\rho_{\rm DE}(0)$, for DESI DR2 BAO $+$
CMB $+$ Union3.1 $+$ CMB lensing. Shaded bands are the $68\%$ and
$95\%$ credible regions; the dash-dotted line marks matter--dark-energy
equality at $z_{\rm DE}=0.33$.}
\end{figure}
\begin{figure*}[!htbp]
\includegraphics[width=0.49\textwidth]{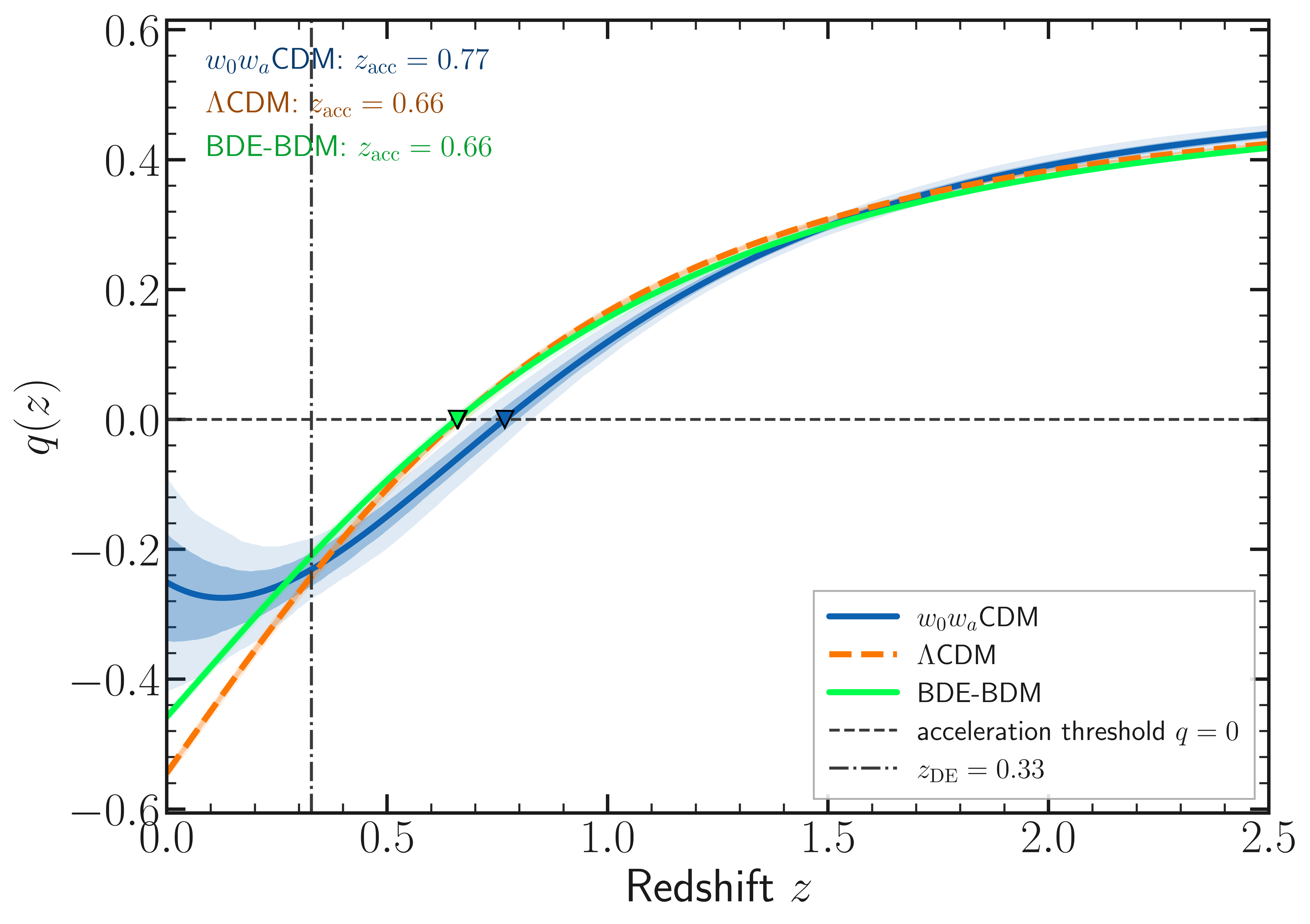}\hfill
\includegraphics[width=0.49\textwidth]{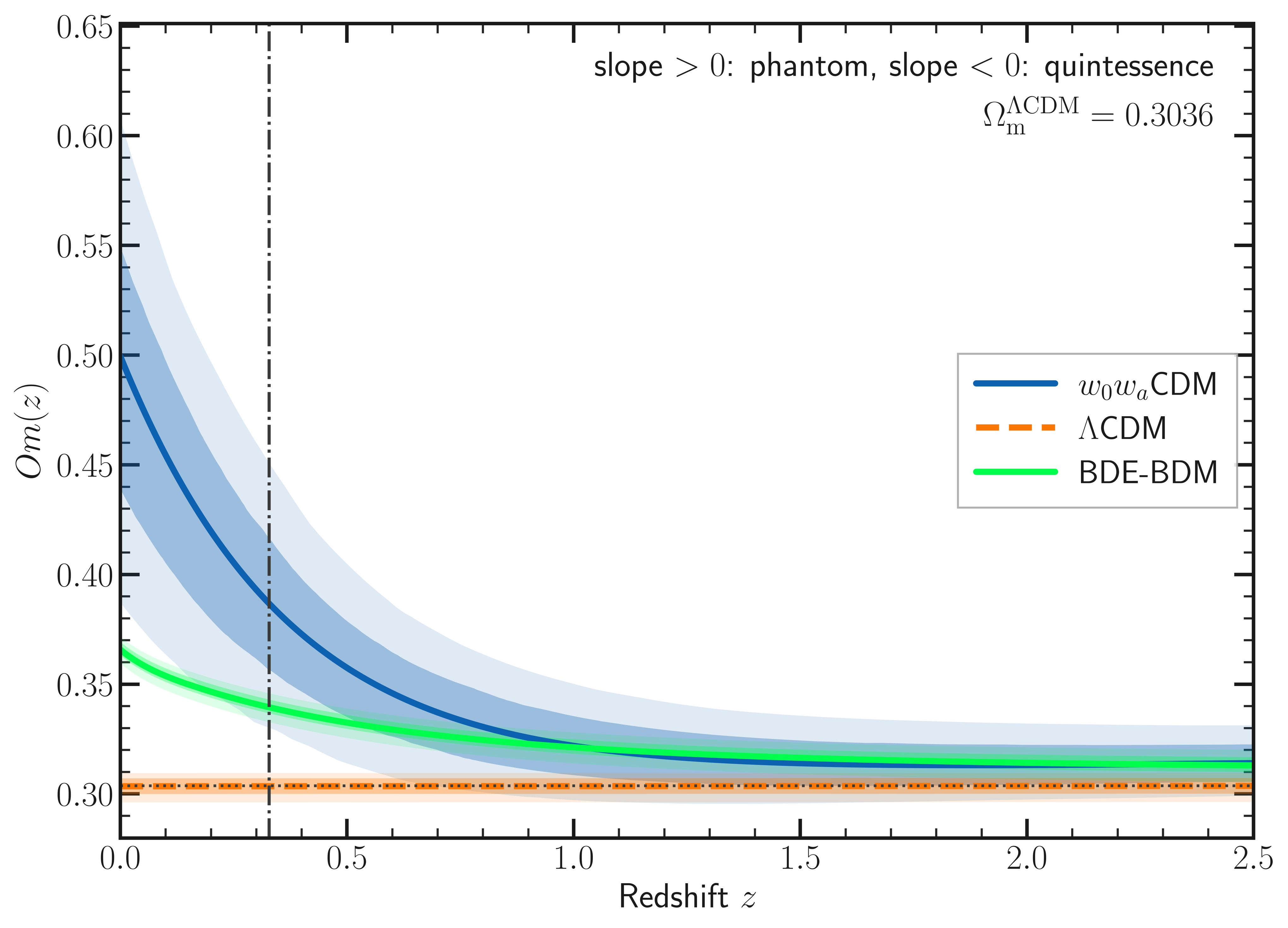}
\caption{\label{fig:qom}%
Parametrization-independent diagnostics for DESI DR2 BAO
$+$ CMB $+$ Union3.1 $+$ CMB lensing data combination.
\textit{Left}: deceleration parameter $q(z)$, with the dashed line at
the acceleration threshold $q=0$ and triangles marking its crossing.
\textit{Right}: the $Om(z)$ diagnostic, which equals \Om\ at all
redshifts for a cosmological constant and has negative slope for
quintessence and positive slope for phantom dark energy; the dotted line
is $\Om^{\Lambda\rm CDM}=0.3036$.}
\end{figure*}
\subsection{The equation of state}
\label{sec:eos}

Figure~\ref{fig:background} shows the equation of state of the three
models at the marginalized means. \LCDM\ is flat at $w=-1$ by
construction. The unified model thaws: having settled at $w\simeq-1$
through the long potential-dominated epoch, it leaves that value at
redshifts of order unity, as described in Sec.~\ref{sec:model}, and rises
monotonically to the present without ever
crossing the phantom divide, so that $w>-1$ holds at every redshift. The
\wowa\ fit runs the other way, lying below $-1$ for $z\gtrsim0.4$ and
climbing steeply only as $a\to1$. The two dynamical descriptions
therefore agree that dark energy departs from a cosmological constant
today, and disagree about the history that brought it there.

The widths differ as sharply as the shapes. The BDE-BDM band is narrower
than the line width at every redshift, while the \wowa\ band widens rapidly
beyond $z\simeq0.5$ and admits both signs of the deviation at high
redshift. The contrast is not evidential: it registers that $w(z)$
follows from the field equations once \Lc\ is fixed, whereas $w_0$ and
$w_a$ are free.

The distinction between the two dynamical descriptions is clearest in
the dark-energy density itself. Figure~\ref{fig:fde} shows
$f_{\rm DE}(z)\equiv\rho_{\rm DE}(z)/\rho_{\rm DE}(0)$. Because BDE-BDM
satisfies $w>-1$ at all times, its density decreases monotonically with
cosmic time, so $f_{\rm DE}$ rises with redshift and flattens near
$1.13$ by $z=2.5$. The \wowa\ fit behaves differently: it is phantom at
$z\gtrsim0.4$, where the dark-energy density grows as the Universe
expands, so $f_{\rm DE}$ reaches a maximum of $1.14$ at $z=0.40$,
crosses unity near $z=1.05$, and falls to $0.62$ by $z=2.5$. The two
models therefore agree closely at $z\lesssim0.5$, where the data are
most constraining, and separate at higher redshift, where a phantom
history requires the dark-energy density to have been smaller in the
past than it is today.

Projected onto the \wowa\ plane following
Refs.~\cite{WolfGarcia2024,WolfGarciaParam2025}, the model occupies a
distinct region of the $(w_0,w_a)$ plane, as the right panel of
Fig.~\ref{fig:background} shows.
Taking $w_{\rm BDE}$ and its first derivative at $a=1$ from the
integrated background, the chains return
\begin{equation}
  w_0 = -0.9262\pm0.0005, \qquad
  w_a = -0.854\pm0.005 ,
  \label{eq:w0waproj}
\end{equation}
For Union3.1, the other two compilations differ by less than $0.001$.
The uncertainty on this point is two orders of magnitude smaller than
the \wowa\ constraint from the same data, so it appears in
that panel as a point rather than a contour.
These errors come from propagating the model parameters through the
field equations; no fit to $(w_0,w_a)$ was performed. One caveat attends
the comparison with the \wowa\ contours.
Equation~\eqref{eq:w0waproj} realizes the first of the three
prescriptions catalogued in
Refs.~\cite{WolfGarcia2024,WolfGarciaParam2025}; fitting the \wowa\ form
over a finite redshift range, or matching $H(z)$ instead, would place
the model at numerically different points of the same plane.

\subsection{Deceleration parameter and \texorpdfstring{$Om$}{Om} diagnostics}
\label{sec:diagnostics}

The comparison in the previous subsection is made at the level of $w(z)$
and therefore presupposes that an equation of state describes the dark energy. Two diagnostics built directly from the expansion rate avoid that assumption and provide a null test of \LCDM\
\cite{Sahni2008,Myrzakulov2023}. Figure~\ref{fig:qom} shows both for the
three models.

The deceleration parameter,
\begin{equation}
  q(z) = -\frac{\ddot a\,a}{\dot a^2},
  \label{eq:qz}
\end{equation}
changes sign at the onset of acceleration. BDE-BDM and \LCDM\ place that
transition at $z_{\rm acc}=0.66$, whereas \wowa\ model places it earlier, at
$z_{\rm acc}=0.77$: its phantom behavior at $z\gtrsim0.4$ raises the
dark-energy density in the past and brings the crossing forward.

The $Om$ diagnostic~\cite{Sahni2008},
\begin{equation}
  Om(z) = \frac{H^2(z)/\Hub^2-1}{(1+z)^3-1},
  \label{eq:omz}
\end{equation}
is constructed so that it returns \Om\ at every redshift if the dark
energy is a cosmological constant, so any variation with redshift is a
departure from \LCDM. For dark energy with an unevolving equation of
state, the slope carries the sign of that departure, negative for
quintessence and positive for phantom~\cite{Sahni2008}. What the slope registers is that $w$ stays above $-1$ over the
range plotted, which Fig.~\ref{fig:background} shows directly. BDE-BDM
has a negative slope throughout, decreasing from $0.365$ at $z=0$
towards the \LCDM\ value at high redshift, and its band is several times
narrower than the \wowa\ band at every redshift. This is the expected
signature of a quintessence field with $w>-1$, obtained here without
fitting $w(z)$.

\subsection{Distances and the expansion rate}
\label{sec:distances}

\begin{figure*}[!htbp]
\includegraphics[width=0.434\textwidth]{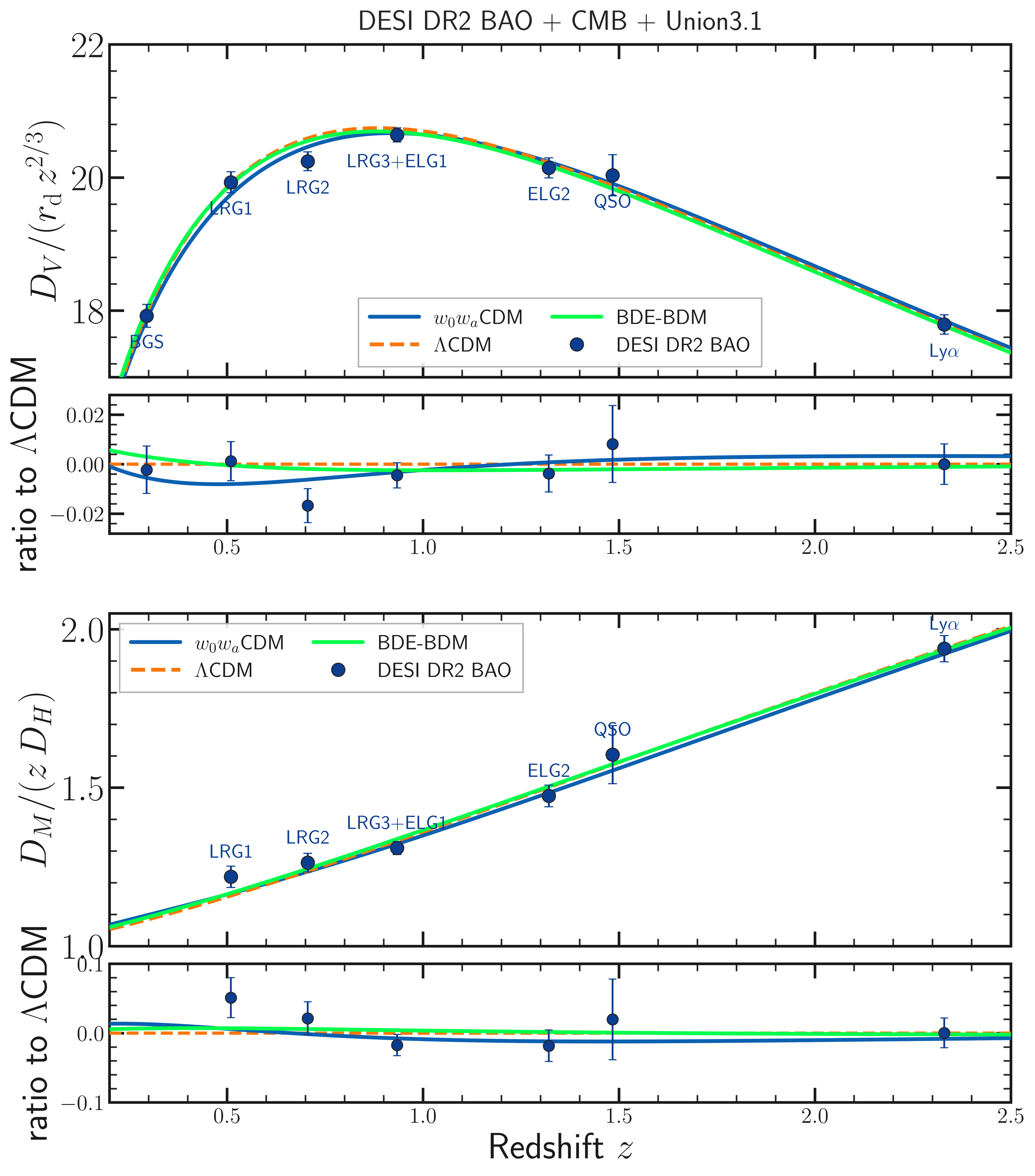}\hfill
\includegraphics[width=0.542\textwidth]{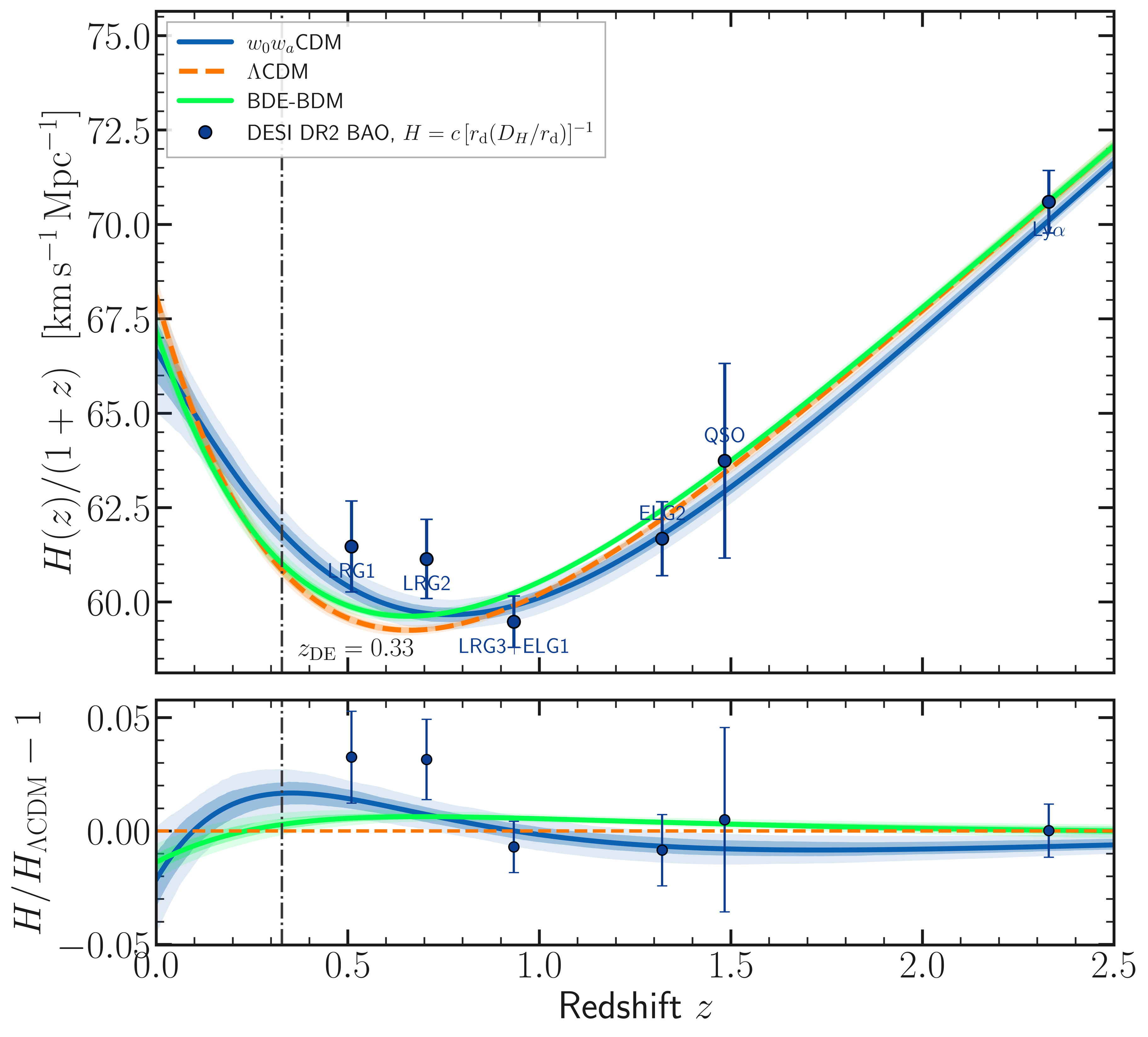}
\caption{\label{fig:bao}%
DESI DR2 BAO measurements against the three models, for DESI DR2 BAO
$+$ CMB $+$ Union3.1 $+$ CMB lensing. \textit{Left}: the
angle-averaged distance $D_V/(r_{\rm d}z^{2/3})$, the factor $z^{-2/3}$
being applied only for clarity, and below it the ratio $D_M/(zD_H)$;
labels identify the DESI tracers. \textit{Right}: the comoving
expansion rate $H(z)/(1+z)$, with the DESI points obtained from
$H=c\,[r_{\rm d}(D_H/r_{\rm d})]^{-1}$ and the dash-dotted line marking
$z_{\rm DE}=0.33$. Sub-panels give the fractional difference from
\LCDM; bands are the $68\%$ and $95\%$ credible regions.}
\end{figure*}
\begin{figure*}[!htbp]
\includegraphics[width=0.98\textwidth]{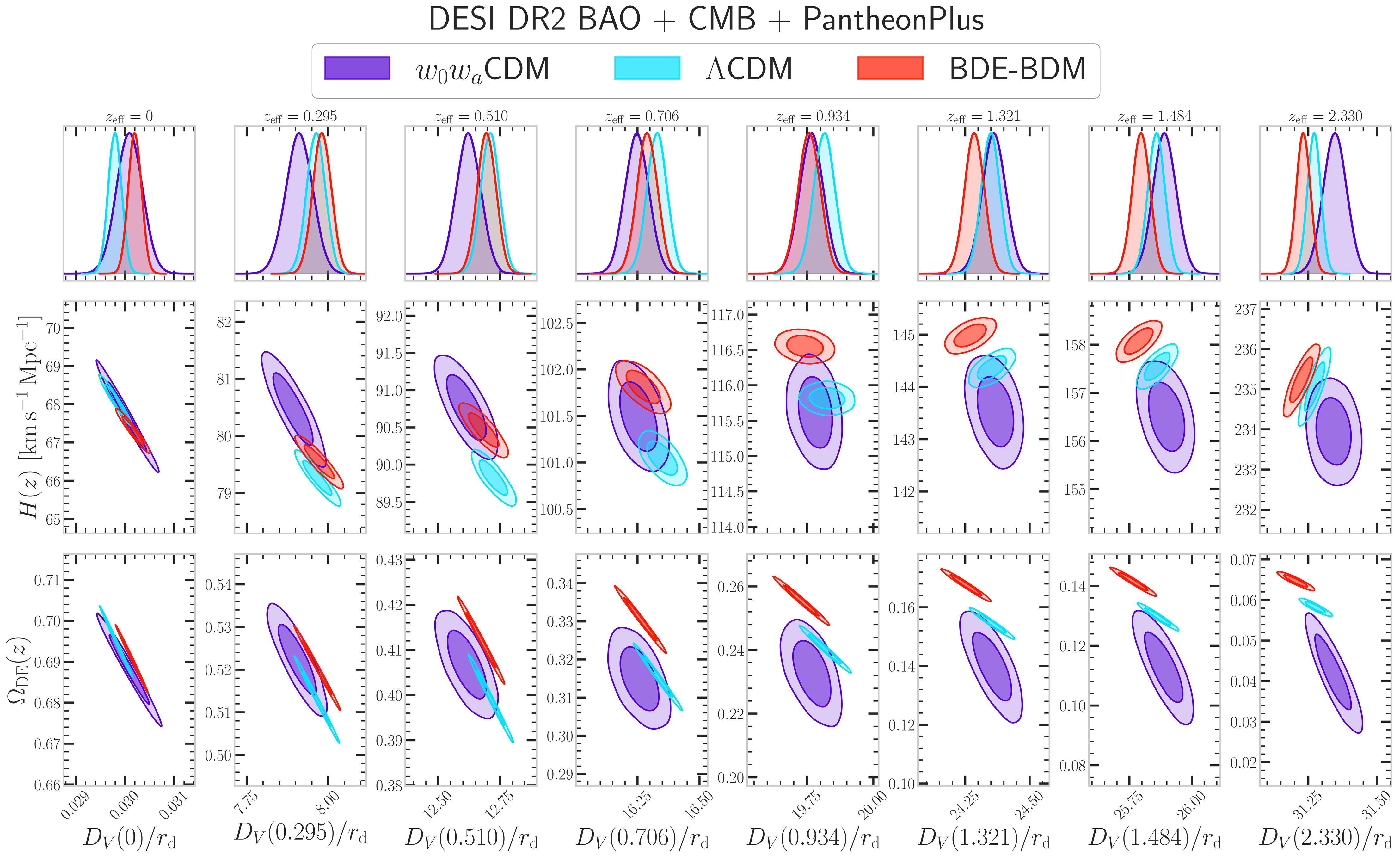}
\caption{\label{fig:dvgrid}%
Constraints at the eight DESI DR2 effective redshifts for DESI DR2 BAO
$+$ CMB $+$ PantheonPlus $+$ CMB lensing. \textit{Top row}:
marginalized posteriors on the angle-averaged distance
$D_V(z)/r_{\rm d}$. \textit{Middle row}: joint $68\%$ and $95\%$
constraints on $D_V(z)/r_{\rm d}$ and the expansion rate $H(z)$.
\textit{Bottom row}: the same against the dark-energy fraction
$\Omega_{\rm DE}(z)$.}
\end{figure*}
The BAO and supernova data constrain the model through the
distance--redshift relation, and it is worth showing directly how well
the predicted expansion history reproduces the DESI DR2 measurements.
The left panel of Fig.~\ref{fig:bao} compares the angle-averaged
distance
$D_V\equiv(zD_M^2D_H)^{1/3}$, with $D_M$ the transverse comoving
distance and $D_H\equiv c/H$, in units of the sound horizon $r_{\rm d}$,
and the ratio $D_M/(zD_H)$, against the seven DESI DR2 tracers. All
three models pass through the measurements, and the lower panels show
that BDE-BDM stays within $0.5\%$ of \LCDM\ across the whole range,
while the \wowa\ fit dips about $1\%$ below it near $z\simeq0.7$. The two
tracers that pull hardest, LRG1 and LRG2, sit above \LCDM\ in $D_M$ and
are matched slightly better by BDE-BDM than by \wowa\ model.
\begin{figure}[!htbp]
\includegraphics[width=\columnwidth]{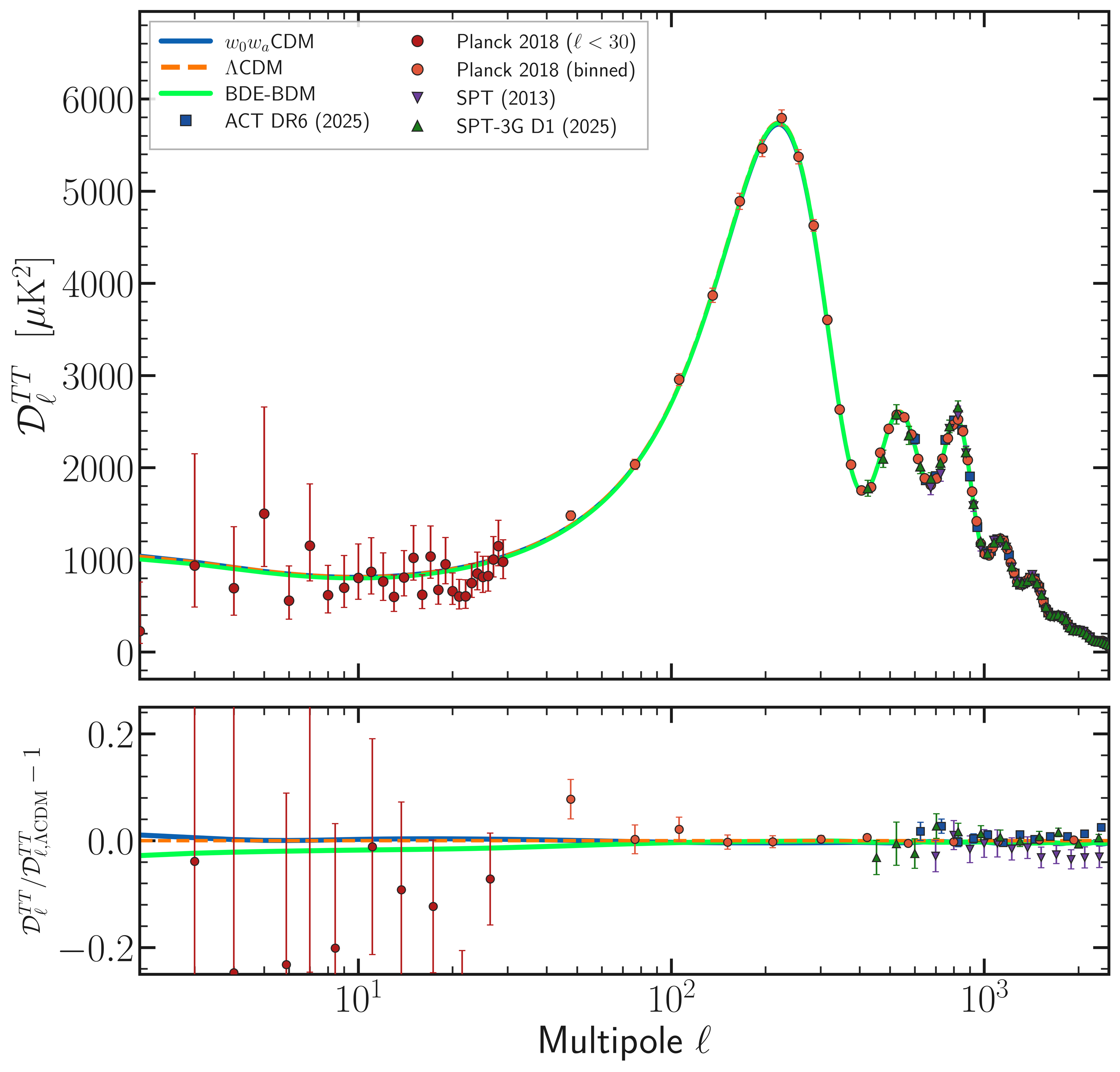}
\caption{\label{fig:cmbtt}%
CMB temperature power spectrum $\mathcal{D}^{TT}_\ell$ at the
marginalized means of DESI DR2 BAO $+$ CMB $+$ Union3.1 $+$ CMB
lensing, with Planck 2018, ACT DR6, SPT and SPT-3G measurements \cite{Planck2020VI, ACTDR6lens}. The
lower panel gives the fractional difference from \LCDM.}
\end{figure}
The right panel shows the same comparison for the expansion rate
$H(z)/(1+z)$, where the DESI points are obtained from the measured
$D_H/r_{\rm d}$. The three histories are indistinguishable at the
precision of the data for $z\gtrsim1$. Below that they separate: \wowa{}
exceeds \LCDM\ by about $1.5\%$ near $z\simeq0.35$ and falls below it at
$z=0$, whereas BDE-BDM rises to at most $0.5\%$ above \LCDM\ near
$z\simeq0.6$ and returns to it by $z\simeq2$. The vertical line marks
matter--dark-energy equality at $z_{\rm DE}=0.33$, close to where the
\wowa\ deviation peaks.

The same information can be displayed in the plane of the observables
the BAO analysis actually reports, at each effective redshift.
Figure~\ref{fig:dvgrid} shows the marginalized posteriors on
$D_V(z)/r_{\rm d}$ at the eight DESI effective redshifts, and their
joint constraints with the expansion rate $H(z)$ and with the
dark-energy fraction $\Omega_{\rm DE}(z)$, for PantheonPlus, the
compilation with the weakest preference for evolving dark energy; the
separation between models seen there is therefore the smallest our data
give. Three features stand out. The BDE-BDM contours are markedly
smaller than the \wowa\ ones at every redshift, because the expansion
history carries no dark-energy parameter beyond \Lc. The orientation of the BDE-BDM
contour in the $D_V$--$H$ plane changes with redshift, following \LCDM\
at low $z$ and moving away from it above $z\simeq0.7$, while the \wowa\
contours keep the fixed orientation set by the $w_0$--$w_a$ degeneracy.
And the dark-energy fraction separates the three models more sharply
than either distance alone: the three predictions for
$\Omega_{\rm DE}(z)$ overlap at $z=0$, as the fit to the same data
requires, and become progressively disjoint towards $z=2.33$, where
BDE-BDM sits above \LCDM\ and \wowa\ below it. This is the quantity in
which the models differ most, and none of the likelihoods used here
constrains it directly.

\subsection{The cosmic microwave background}
\label{sec:cmbresult}

Figure~\ref{fig:cmbtt} shows the CMB temperature spectrum. The
three models are visually identical at the resolution of the data, as
they must be for models sharing the same pre-recombination physics. The
lower panel shows that BDE-BDM sits a few percent below \LCDM\ at
$\ell\lesssim10$, where cosmic variance is large, and converges to it
across the acoustic peaks and the damping tail. This is the residual
imprint of the transient extra radiation \Next\ present before
condensation, and it is the same effect that produces the CamSpec
penalty discussed in Sec.~\ref{sec:comparison}.

\subsection{Growth rate and structure formation}
\label{sec:pkresult}

\begin{figure}[!htbp]
\includegraphics[width=\columnwidth]{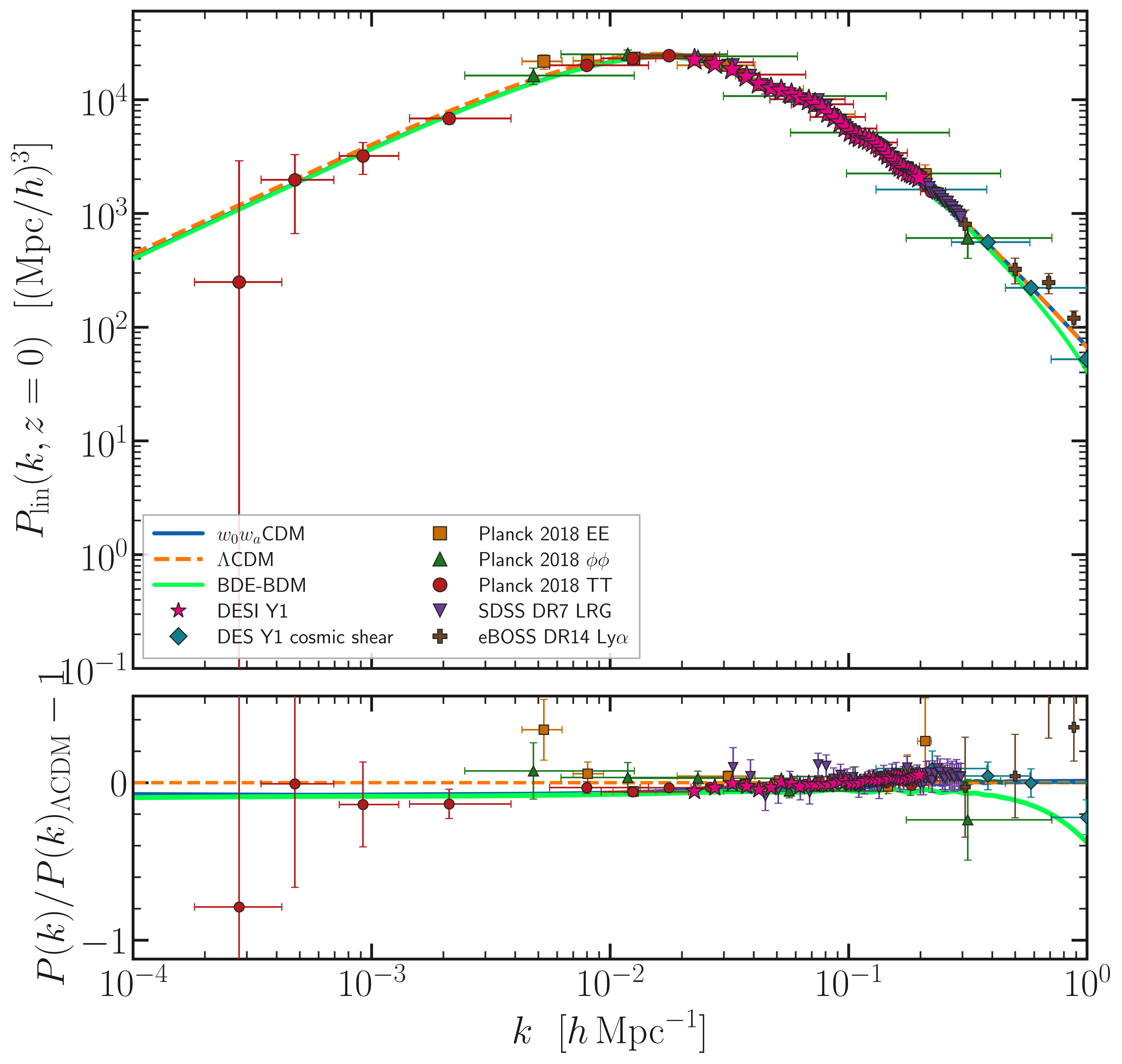}
\caption{\label{fig:pk}%
\textit{Top}: the $z=0$ linear matter power spectra of the three models,
at the marginalized means of DESI DR2 BAO $+$ CMB $+$ Union3.1 $+$ CMB
lensing dataset combination. Overplotted
along $P(k)$ are data measurements from Planck 2018 TT, EE and $\phi\phi$, SDSS DR7
LRG, DES Y1 cosmic shear, eBOSS DR14 Lyman-$\alpha$ given in \cite{Chabanier2019} and DESI Y1 \cite{Cereskaite_2026}. \textit{Bottom}: the fractional departure from
\LCDM.}
\end{figure}

\begin{figure*}[!htbp]
\includegraphics[width=0.88\textwidth]{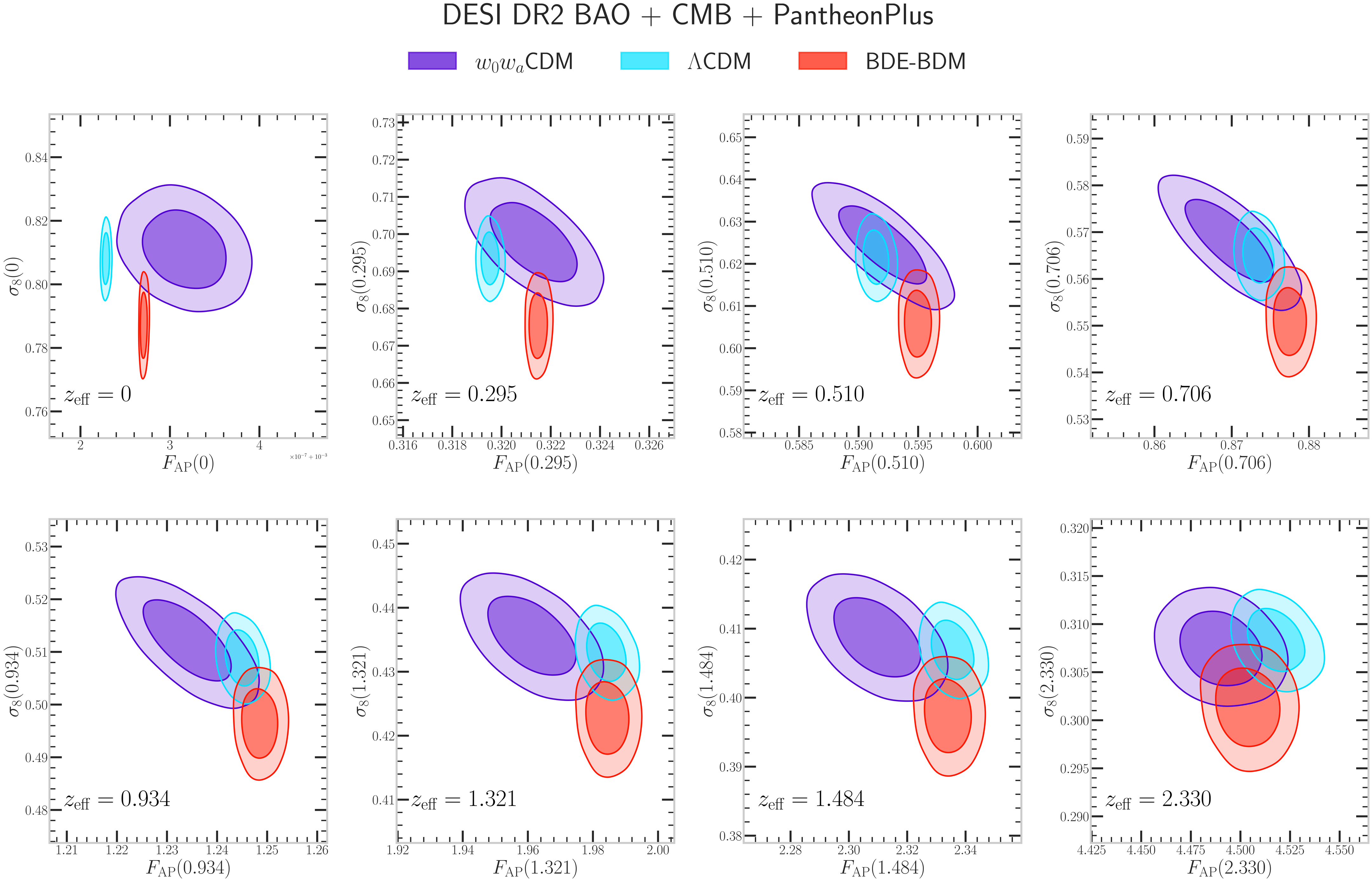}
\caption{\label{fig:apgrid}%
Joint $68\%$ and $95\%$ constraints on the Alcock--Paczynski parameter
$F_{\rm AP}(z)=D_M(z)H(z)/c$ and the amplitude of structure $\sig(z)$
at the eight DESI DR2 effective redshifts, for DESI DR2 BAO $+$ CMB $+$
PantheonPlus $+$ CMB lensing.}
\end{figure*}

\begin{figure}[!htbp]
\includegraphics[width=\columnwidth]{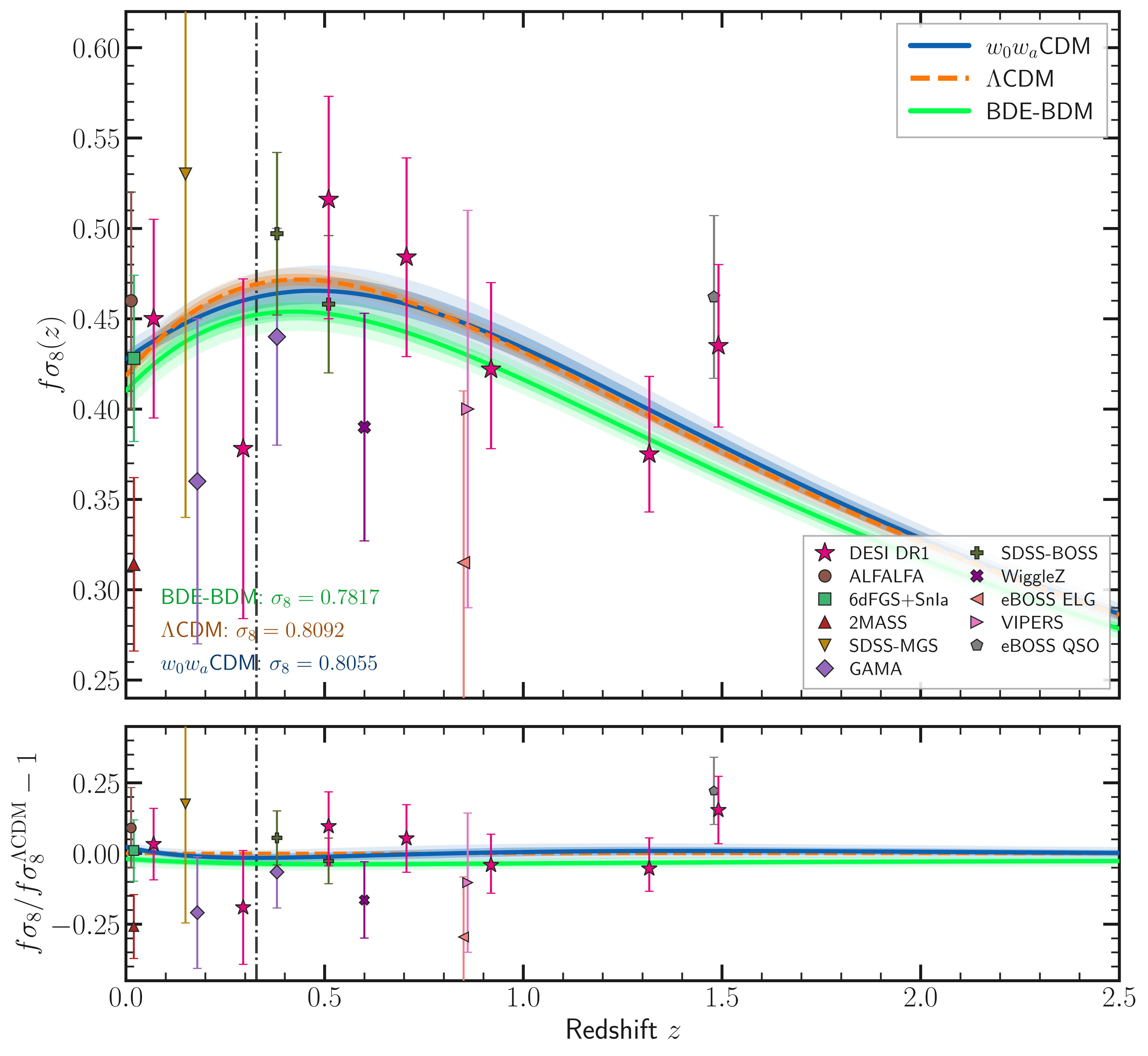}
\caption{\label{fig:fsigma8}%
The product $f\sig(z)$ of the growth rate and the amplitude of structure
for the three models at the marginalized means of
DESI DR2 BAO $+$ CMB $+$ Union3.1 $+$ CMB lensing, with bands giving
the $68\%$ and $95\%$ credible regions, against redshift-space
distortion measurements from DESI DR1 and earlier surveys. The
dash-dotted line marks $z_{\rm DE}=0.33$ and the lower panel gives the
fractional difference from \LCDM.}
\end{figure}

The competition between the two dark-sector components described in
Sec.~\ref{sec:smallscales} is resolved, for the preferred parameters, in
favor of the free-streaming suppression, and Fig.~\ref{fig:pk} shows
the outcome. The half-suppression scale defined in Eq.~\eqref{eq:khalf}
takes the value $\khalf=1.16\,h\,\mathrm{Mpc}^{-1}$ at the best fit,
with the comoving horizon at the transition close beside it at
$k_{\rm hor}(\ac)\simeq1.4\,h\,\mathrm{Mpc}^{-1}$, so that the two
scales of Sec.~\ref{sec:smallscales} act together as anticipated.

At the marginalized means the BDE-BDM spectrum lies
$6.0\%$ below the matched \LCDM\ one at $k=0.05\,h\,\mathrm{Mpc}^{-1}$
and $5.0\%$ below at $k=0.2\,h\,\mathrm{Mpc}^{-1}$ (Union3.1; the other
two compilations agree to $0.1\%$). Neither of these is a free-streaming
effect: with $\khalf=1.16\,h\,\mathrm{Mpc}^{-1}$ the suppression is
still negligible there, and the offset is the broadband amplitude
difference that follows from the lower \sig\ preferred by the unified
model, $0.782$ against $0.809$ for \LCDM. Free streaming becomes the
dominant effect only beyond $k\simeq0.3\,h\,\mathrm{Mpc}^{-1}$, the
deviation growing to $-11\%$ at $0.5$, $-38\%$ at $1.0$, and $-73\%$ at
$1.5\,h\,\mathrm{Mpc}^{-1}$. For comparison, \wowa\ departs from \LCDM\
by less than $1\%$ over the whole of this range, so the small-scale
spectrum separates the unified model from both alternatives more sharply
than any background observable. Integrated over the standard window,
these deviations amount to a reduction of \sig\ by $0.027$ with respect
to \LCDM\ for the same data, which is the origin of the structure
constraints examined next.

\subsection{Amplitude of structure and the bound velocity}
\label{sec:vcresult}

The amplitude of structure is systematically lower than in \LCDM:
\begin{align}
  \sig    &= 0.782\pm0.008, \qquad
  \Seight  = 0.796\pm0.010, \nonumber
\end{align}
against $\sig=0.809\pm0.005$ and $\Seight=0.814\pm0.007$ for \LCDM\ with
the same data, and the same pattern holds for the other two SNe
compilations (Table~\ref{tab:full}). As we show below, this reduction
has two distinct origins with very different degrees of robustness, and
we therefore analyze it together with the constraint on \vc.

The deficit is visible directly in the growth rate.
Figure~\ref{fig:fsigma8} shows $f\sig(z)$, with $f\equiv d\ln\delta_m/d\ln a$
the linear growth rate of the matter density contrast $\delta_m$, for the
three models against the redshift-space distortion measurements compiled
from DESI DR1 and
earlier surveys. BDE-BDM lies below \LCDM\ across the whole range, by
$3$--$5\%$, and the offset is close to constant in redshift: it is
inherited from the lower present-day normalization rather than generated
by a difference in the growth history. The \wowa\ fit, by contrast, tracks
\LCDM\ to better than $1\%$ over $0<z<2.5$ despite its two extra
parameters, so the three models are separated in this observable
essentially by \sig\ alone. The scatter of the measurements is larger
than the separation between the curves, and no useful discrimination
follows from the present compilation; the figure is shown because the
sign and the redshift dependence of the offset are predictions that
Stage-IV redshift surveys will test.

We can separate the same deficit from the geometry.
Figure~\ref{fig:apgrid} pairs a purely geometrical quantity with a
growth quantity at the eight DESI DR2 effective redshifts. The
Alcock--Paczynski parameter $F_{\rm AP}(z)\equiv D_M(z)H(z)/c$ is what
the anisotropy of the clustering measures, independently of the
sound-horizon calibration, and $\sig(z)$ is the amplitude of structure
at the same epoch. The BDE-BDM constraint on $F_{\rm AP}$ is as tight as
that of \LCDM\ and much tighter than that of \wowa\ at every redshift,
because the expansion history has no free shape. The displacement
between the models is therefore almost entirely along the $\sig$
direction, and it is largest at $z=0$, where the BDE-BDM contour lies
below both others, and closes towards $z\simeq2.3$, where the three
overlap. Together with Fig.~\ref{fig:fsigma8}, this places the lower
amplitude of structure in the late-time normalization rather than in the
geometry, and it is the reason the deficit is worth analyzing alongside
\vc\ rather than as a feature of the expansion history.

For the velocity, we obtain
\begin{equation}
  \vc = 0.463\pm0.171 ,
  \qquad
  0.10 < \vc < 1/\sqrt2 \ \ (95\%~\mathrm{C.L.}),
  \label{eq:vcfit}
\end{equation}
The upper end of the $95\%$ interval rests on the prior boundary, so
that the velocity is bounded from below by the data and from above only
by the prior. The posterior is strongly asymmetric: its $68\%$ interval
is $[0.41,0.71]$ rather than the symmetric range implied by the quoted
standard deviation, and the mean lies well below the peak. How far the
cold limit is disfavored also depends on the supernova sample. With
PantheonPlus and Union3.1 the $95\%$ interval is bounded from below, at
$\vc>0.13$ and $\vc>0.10$, whereas with DES-Dovekie it reaches the lower
prior edge and the velocity is unconstrained.
So wide an interval is what one should anticipate here. Scanning $(\ac,\vc)$ against CMB, SNe~Ia, and BAO data,
Ref.~\cite{Mastache2020} was able to bound \ac\ from above but obtained
no restriction on \vc\ at all, the reason being simply that the velocity
manifests itself on scales shorter than those such probes reach. The
combination used here is of the same type, its only addition being CMB
lensing, which acts above the cutoff.

The scales at which the velocity acts are absent from every likelihood
we use, and the free-streaming integral Eq.~\eqref{eq:lambdafs}
identifies them. Evaluating it over the best-fit
background gives $\lambda_{\rm fs}=2.0\,h^{-1}\,\mathrm{Mpc}$ at
$\vc=0.5$, hence a free-streaming wavenumber
$k_{\rm fs}=2\pi/\lambda_{\rm fs}=3.1\,h\,\mathrm{Mpc}^{-1}$, consistent
with the half-suppression scale of Sec.~\ref{sec:pkresult} obtained
from the full transfer function, given the usual factor of two to three
between the two definitions. The associated half-mode mass,
\begin{equation}
  M_{1/2} = \frac{4\pi}{3}\,\bar\rho_m
            \left(\frac{\pi}{\khalf}\right)^{3}
          \simeq 7\times10^{12}\,M_\odot\,h^{-1},
  \label{eq:Mfs}
\end{equation}
with $\bar\rho_m$ the present mean matter density, exceeds the mass of a
Milky Way halo, and a thermal relic with the same
half-mode scale would have a mass of order $0.2$~keV; inverting the same
integral, the free-streaming scale matches the half-mode wavenumber of a
$3$~keV ($5.3$~keV) relic for $\vc\simeq0.08$ ($\simeq0.04$). However,
future observations at smaller scales can restrict \vc\ further; the
dependence of our results on the velocity is quantified in
Sec.~\ref{sec:coldlimit}.

\subsection{Results independent of the bound velocity}
\label{sec:coldlimit}

Because the data leave \vc\ largely open, we separate the results that
depend on it from those that do not. Table~\ref{tab:coldlimit} compares
the full marginalized constraints with those of the subsample of the
chains with $\vc<0.1$, which carries $1.8\%$ of the posterior weight and
has a mean velocity of $0.055$. Tightening the cut to $\vc<0.05$ changes
no entry by more than $0.0004$, so the subsample represents the cold
limit.

The condensation scale, the transition epoch, the expansion rate and the
matter density do not depend on \vc: their correlation coefficients with
it are below $0.08$ in magnitude, and between the full posterior and the
cold limit \Lc\ moves from $43.67\pm0.13$ to $43.63\pm0.13$~eV, \Hub\
from $67.11\pm0.26$ to $67.02\pm0.25$, and \Om\ from $0.3111\pm0.0036$
to $0.3125\pm0.0036$, in each case by less than half a standard
deviation. The velocity enters the dark-energy sector only through the
energy partition: the Lorentz factor $\gamma_c$ in
Eq.~\eqref{eq:rhobdmpre} raises the share taken by the massive states,
so \fbde\ grows from $0.936\pm0.006$ to $0.9445\pm0.0005$ in the cold
limit, and $w_0$ and $w_a$ follow, to $-0.92682\pm0.00031$ and
$-0.8461\pm0.0034$. These shifts exceed the marginalized uncertainties
because those uncertainties are themselves set by \vc; in absolute
terms they amount to $0.07\%$ in $w_0$ and $1\%$ in $w_a$, far below the
resolution of any expansion-history probe, so the predicted $w(z)$ of
Sec.~\ref{sec:background} is the same in either case.

The quality of the fit does depend on \vc. The supernova term is
independent of it, but towards the cold limit the DESI BAO $\chi^2$ rises
by $1.0$--$1.5$ and the low-$\ell$ EE $\chi^2$ by $1.3$--$2.2$ across the
three compilations. The first follows the smaller $|w_a|$ that
accompanies the larger \fbde, the second the higher optical depth,
$\tau_{\rm reio}=0.063\pm0.008$ against $0.057$, that accompanies a
higher $A_s$ at fixed $A_se^{-2\tau_{\rm reio}}$. The cold limit
therefore costs $\Delta\chi^2=2.3$--$3.6$ at the best-fit minimum
(Sec.~\ref{sec:vcresult}) and lowers the Bayesian evidence. Restricting
the prior to $\vc<0.1$ changes $\ln\mathcal{Z}$ by the logarithm of the
ratio of posterior to prior mass in that range, $-1.30$, $-2.07$ and
$-2.18$ for PantheonPlus, Union3.1 and DES-Dovekie, so that
$\ln B_{\mathrm{BDE\text{-}BDM},\Lambda\mathrm{CDM}}$ becomes $2.79$,
$2.05$ and $2.65$: in the cold limit BDE-BDM remains favored over
\LCDM, at the weak-to-moderate level.

The amplitude of structure separates into two parts. In the cold limit
$\sig=0.797\pm0.006$, below the \LCDM\ value $0.809\pm0.005$ by $0.012$.
The velocity removes a further $0.015$, bringing \sig\ to
$0.782\pm0.008$; about two-thirds of this second step is the
free-streaming suppression at fixed $A_s$ and one-third the accompanying
decrease of $A_s$. Of the total reduction relative to \LCDM, $44\%$ is
therefore independent of \vc\ and $56\%$ depends on it. For \Seight\ the
separation is complete: the cold-limit value $\Seight=0.814\pm0.007$
coincides with that of \LCDM, the larger \Om\ compensating the lower
\sig, so the entire reduction to $\Seight=0.796\pm0.010$ depends on the
velocity.

However, future observations at smaller scales can restrict \vc\
further. Should they confine it to the cold limit, the third column of
Table~\ref{tab:coldlimit} applies, and the unified model reduces to bound
dark energy with a dark-matter component that is cold at condensation,
retaining a distinctive small-scale signature in the residual suppression
near $k\sim10\,h\,\mathrm{Mpc}^{-1}$ and in the BDE-driven enhancement
discussed in Secs.~\ref{sec:smallscales} and~\ref{sec:pkresult}. The
free-streaming scale at the best fit, $\khalf=1.16\,h\,\mathrm{Mpc}^{-1}$,
lies in the range accessible to such observations
(Sec.~\ref{sec:discussion}), which can therefore measure \vc\ directly.

\begin{table}[t]
\caption{\label{tab:coldlimit}%
Dependence of the BDE-BDM constraints on the bound velocity, for
BAO+CMB+Union3.1. The second column marginalizes over the full \vc\
prior; the third restricts the chains to $\vc<0.1$, which represents the
cold limit. The \LCDM\ values for the same data are shown for reference;
$\ln B$ is the Bayes factor against \LCDM.}
\begin{ruledtabular}
\scriptsize
\setlength{\tabcolsep}{3pt}
\begin{tabular}{l c c c}
 & BDE-BDM & BDE-BDM & \LCDM \\
 & (marginal) & ($\vc<0.1$) & \\
\colrule
$\Lc$~[eV]     & $43.67 \pm 0.13$       & $43.63 \pm 0.13$       & -- \\
$10^{6}\,\ac$  & $2.5052 \pm 0.0077$    & $2.5075 \pm 0.0076$    & -- \\
$\fbde$        & $0.9357 \pm 0.0061$    & $0.9445 \pm 0.0005$    & -- \\
$w_0$          & $-0.92615 \pm 0.00048$ & $-0.92682 \pm 0.00031$ & $-1$ \\
$w_a$          & $-0.8535 \pm 0.0053$   & $-0.8461 \pm 0.0034$   & $0$ \\
$\Hub$         & $67.11 \pm 0.26$       & $67.02 \pm 0.25$       & $68.12 \pm 0.28$ \\
$\Om$          & $0.3111 \pm 0.0036$    & $0.3125 \pm 0.0036$    & $0.3036 \pm 0.0035$ \\
$\sig$         & $0.782 \pm 0.008$      & $0.797 \pm 0.006$      & $0.809 \pm 0.005$ \\
$\Seight$      & $0.796 \pm 0.010$      & $0.814 \pm 0.007$      & $0.814 \pm 0.007$ \\
$\ln B$        & $4.12$                 & $2.05$                 & $0$ \\
\end{tabular}
\end{ruledtabular}
\end{table}

Two key features of Table~\ref{tab:full} are worth stating explicitly.
The values of \Lc, \ac, \fbde, $w_0$, $w_a$, \sig, \Seight, and \vc\ are
stable across the three SNe compilations, whereas the \wowa\ values of
$w_0$ and $w_a$ shift by more than one standard deviation between them.
The dark sector inferred for the unified model is thus the same whichever
supernova sample is used.
Throughout, the sound-horizon scale $r_{\rm d}$ is nearly identical
among the three models, confirming that the pre-recombination physics is
unchanged and that the differences reside at late-times.

\subsection{Model comparison}
\label{sec:comparison}
\begin{table*}[t]
\caption{\label{tab:dic_full}%
Posterior-mean deviance $\overline{D(\theta)}$, deviance
$D(\overline{\theta})$, penalty term $p_D=\overline{D(\theta)}-D(\overline{\theta})$,
and $\mathrm{DIC}=\overline{D(\theta)}+p_D$ for each model and dataset
combination, together with the differences
$\Delta\chi^2\equiv\chi^2_{\rm model}-\chi^2_{\Lambda\rm CDM}$ and
$\Delta\mathrm{DIC}\equiv\mathrm{DIC}_{\rm model}-
\mathrm{DIC}_{\Lambda\rm CDM}$ relative to the \LCDM\ baseline. All
deviations are computed from the data likelihoods only. Negative values
of $\Delta\chi^2$ or $\Delta\mathrm{DIC}$ favor the alternative model
over \LCDM. The last two columns give the Bayes factor against the
\LCDM\ baseline, $\ln B_{X,\Lambda\rm CDM}$, positive values favoring
the model. The Bayes factor
between any two models follows by subtraction, for example
$\ln B_{\mathrm{BDE\text{-}BDM},\wowa}=+6.21$, $+4.38$ and $+5.39$.}
\begin{ruledtabular}
\begin{tabular}{l l d d d d d d d l}
Dataset & Model &
  \multicolumn{1}{c}{$\overline{D(\theta)}$} &
  \multicolumn{1}{c}{$D(\overline{\theta})$} &
  \multicolumn{1}{c}{$p_D$} &
  \multicolumn{1}{c}{DIC} &
  \multicolumn{1}{c}{$\Delta\chi^2$} &
  \multicolumn{1}{c}{$\Delta\mathrm{DIC}$} &
  \multicolumn{1}{c}{$\ln B$} &
  Evidence \\
\colrule
\multirow{6}{*}{BAO+CMB+PantheonPlus}
 & \LCDM   & 12418.92 & 12404.37 & 14.55 & 12433.47 &   0.00 &   0.00 &  0.00 & reference \\
 & \wowa   & 12410.83 & 12395.04 & 15.80 & 12426.63 &  -9.33 &  -6.84 & -2.11 & none      \\
 & BDE-CDM & 12417.05 & 12402.39 & 14.65 & 12431.70 &  -1.98 &  -1.77 &  1.35 & weak      \\
 & \LBDM   & 12415.41 & 12399.75 & 15.67 & 12431.08 &  -4.62 &  -2.39 &  2.49 & weak      \\
 & \wowaBDM & 12410.43 & 12394.27 & 16.16 & 12426.58 & -10.10 &  -6.89 & -0.82 & none      \\
 & BDE-BDM & 12412.14 & 12399.08 & 13.06 & 12425.19 &  -5.29 &  -8.28 &  4.09 & moderate  \\[2pt]
\multirow{6}{*}{BAO+CMB+Union3.1}
 & \LCDM   & 11043.90 & 11028.55 & 15.35 & 11059.25 &   0.00 &   0.00 &  0.00 & reference \\
 & \wowa   & 11032.98 & 11018.34 & 14.64 & 11047.62 & -10.21 & -11.63 & -0.26 & none      \\
 & BDE-CDM & 11042.35 & 11027.48 & 14.87 & 11057.22 &  -1.07 &  -2.03 &  2.26 & weak      \\
 & \LBDM   & 11040.06 & 11024.27 & 15.79 & 11055.86 &  -4.28 &  -3.39 &  2.51 & moderate  \\
 & \wowaBDM & 11032.81 & 11016.72 & 16.09 & 11048.90 & -11.83 & -10.35 &  1.05 & weak      \\
 & BDE-BDM & 11037.25 & 11022.76 & 14.49 & 11051.73 &  -5.79 &  -7.52 &  4.12 & moderate  \\[2pt]
\multirow{6}{*}{BAO+CMB+DES-Dovekie}
 & \LCDM   & 12648.16 & 12634.58 & 13.58 & 12661.73 &   0.00 &   0.00 &  0.00 & reference \\
 & \wowa   & 12636.80 & 12621.63 & 15.17 & 12651.97 & -12.94 &  -9.76 & -0.56 & none      \\
 & BDE-CDM & 12645.03 & 12629.87 & 15.15 & 12660.18 &  -4.70 &  -1.56 &  2.08 & weak      \\
 & \LBDM   & 12644.87 & 12629.75 & 15.12 & 12659.99 &  -4.83 &  -1.75 &  2.06 & weak      \\
 & \wowaBDM & 12636.35 & 12621.15 & 15.20 & 12651.55 & -13.42 & -10.19 &  0.62 & none      \\
 & BDE-BDM & 12639.83 & 12625.40 & 14.43 & 12654.26 &  -9.18 &  -7.47 &  4.83 & moderate  \\
\end{tabular}
\end{ruledtabular}
\end{table*}

\begin{table*}[t]
\caption{\label{tab:chi2_all}%
Best-fit $\chi^2$ of each data component for \LCDM, and the shift
$\Delta\chi^2$ of each alternative model relative to it (negative values
favor the model), for the three dataset combinations. The totals give the
best-fit deviances $D(\overline{\theta}))$ entering Eq.~\eqref{eq:dic_def} and
reported in Table~\ref{tab:dic_full}. Each block is BAO+CMB with the supernova compilation
named, to which the SNe row refers.}
\begin{ruledtabular}
\begin{tabular}{l l d d d d d d}
 & & \multicolumn{1}{c}{$\chi^2$} &
     \multicolumn{5}{c}{$\Delta\chi^2$ vs \LCDM} \\
\cline{3-3}\cline{4-8}
SNe & Component &
  \multicolumn{1}{c}{\LCDM} &
  \multicolumn{1}{c}{\wowa} &
  \multicolumn{1}{c}{BDE-CDM} &
  \multicolumn{1}{c}{\LBDM} &
  \multicolumn{1}{c}{\wowaBDM} &
  \multicolumn{1}{c}{BDE-BDM} \\
\colrule
\multirow{7}{*}{PantheonPlus}
 & DESI DR2 BAO          & 13.25 & -3.01 & -1.61 & -2.19 & -3.02 & -3.32 \\
 & Planck low-$\ell$ TT  & 22.64 & 0.61 & -0.77 & -0.89 & 0.58 & -1.14 \\
 & Planck low-$\ell$ EE  & 397.35 & -1.39 & -0.43 & -1.50 & -1.51 & -0.98 \\
 & CamSpec TTTEEE        & 10545.59 & -3.24 & 1.81 & -0.20 & -3.12 & 3.00 \\
 & CMB lensing       & 20.14 & -0.07 & 1.88 & -0.58 & -0.65 & 0.03 \\
 & PantheonPlus SNe      & 1405.40 & -2.22 & -2.86 & 0.75 & -2.37 & -2.88 \\
 & \textit{Total}        & 12404.37 & -9.33 & -1.98 & -4.62 & -10.10 & -5.29 \\[2pt]
\multirow{7}{*}{Union3.1}
 & DESI DR2 BAO          & 13.26 & -4.47 & -0.66 & -2.26 & -4.01 & -2.95 \\
 & Planck low-$\ell$ TT  & 22.43 & 0.75 & -0.26 & -0.70 & 0.51 & -0.74 \\
 & Planck low-$\ell$ EE  & 395.88 & -0.09 & 1.77 & -0.14 & 0.13 & -0.17 \\
 & CamSpec TTTEEE        & 10546.61 & -2.63 & -0.99 & -1.37 & -3.88 & 0.64 \\
 & CMB lensing       & 20.01 & -0.32 & 1.46 & -0.33 & -0.63 & -0.04 \\
 & Union3.1 SNe          & 30.36 & -3.46 & -2.39 & 0.52 & -3.95 & -2.53 \\
 & \textit{Total}        & 11028.55 & -10.21 & -1.07 & -4.28 & -11.83 & -5.79 \\[2pt]
\multirow{7}{*}{DES-Dovekie}
 & DESI DR2 BAO          & 13.79 & -3.37 & -2.08 & -2.22 & -3.57 & -3.67 \\
 & Planck low-$\ell$ TT  & 22.64 & 0.62 & -0.50 & -0.78 & 0.69 & -1.03 \\
 & Planck low-$\ell$ EE  & 397.22 & -1.21 & 1.25 & -0.04 & -1.41 & -1.14 \\
 & CamSpec TTTEEE        & 10545.99 & -3.20 & -0.10 & -1.46 & -3.79 & 1.36 \\
 & CMB lensing       & 20.45 & -0.65 & 0.69 & -1.15 & -0.50 & -0.62 \\
 & DES-Dovekie SNe       & 1634.48 & -5.13 & -3.96 & 0.82 & -4.84 & -4.07 \\
 & \textit{Total}        & 12634.58 & -12.94 & -4.70 & -4.83 & -13.42 & -9.18 \\
\end{tabular}
\end{ruledtabular}
\end{table*}
We compare the models with three statistics. The first is the
difference in best-fit effective chi-square,
\begin{equation}
  \Delta\chi^2 \equiv \chi^2_{\rm model} - \chi^2_{\Lambda\rm CDM},
  \label{eq:dchi2}
\end{equation}
negative when the model accounts for the data better than the baseline.
The second is the deviance information
criterion~\cite{Spiegelhalter2002,Liddle2007},
\begin{equation}
  \mathrm{DIC} \equiv D(\overline{\theta}) + 2\,p_D
               = \overline{D(\theta)} + p_D,
  \qquad
  p_D \equiv \overline{D(\theta)} - D(\overline{\theta}),
  \label{eq:dic_def}
\end{equation}
built from the deviance $D(\overline{\theta})=-2\ln\mathcal{L}(\theta)$, its
posterior average, and its value at the best-fit point $\overline{\theta}$. Here $p_D$ counts the
parameter directions the data actually constrain and supplies the
complexity penalty. Since $D(\overline{\theta})=\chi^2_{\min}$, the two
statistics obey
\begin{equation}
  \Delta\mathrm{DIC} - \Delta\chi^2
  = 2\bigl(p_D^{\rm model} - p_D^{\Lambda\rm CDM}\bigr),
  \label{eq:dic_chi2_identity}
\end{equation}
so anything separating them comes from the effective parameter counts
alone. For thresholds we use the
scale of the DESI extended dark-energy study~\cite{Lodha2025}: below $2$
carries no weight, $2$ to $5$ is a positive preference, $5$ to $10$
strong, above $10$ decisive.

Neither statistic accounts for the prior volume a model spends to reach
its best fit, so we also compute the Bayesian evidence
\begin{equation}
\begin{split}
  \mathcal{Z} &\equiv \int \mathcal{L}(D|\theta,M)\,\pi(\theta|M)\,d\theta ,\\
  \ln B_{X,Y} &\equiv \ln\mathcal{Z}_X - \ln\mathcal{Z}_Y ,
\end{split}
  \label{eq:evidence}
\end{equation}
positive values favoring $X$, evaluated from the chains with the
estimator of Ref.~\cite{Heavens2017} as in Appendix~A of
Ref.~\cite{Lodha2025}, with an uncertainty of $\simeq0.4$. Here we use
the Jeffreys scale in Trotta's form~\cite{Trotta2008}: $\ln B>1$ weak,
$>2.5$ moderate, $>5$ strong, a different scale from the
$\Delta\mathrm{DIC}$. Deviances use the data likelihoods
alone. We do not
quote the Akaike criterion, whose penalty would count \Lc\ exactly as it
counts $w_0$ and $w_a$ even though Eq.~\eqref{eq:Lc} imposes the range
\Lc\ explores.

Table~\ref{tab:dic_full} gives
$\overline{D(\theta)}$, $D(\overline{\theta})$, $p_D$, DIC, $\Delta\chi^2$ and
$\Delta\mathrm{DIC}$; Table~\ref{tab:chi2_all} the per-likelihood
decomposition. Both dynamical models fit every combination better than
\LCDM. With its two extra parameters, \wowa\ lowers the best-fit $\chi^2$ by
$9.33$ (PantheonPlus), $10.21$ (Union3.1) and $12.94$ (DES-Dovekie);
BDE-BDM, which samples one parameter fewer and adjusts none in the
dark-energy sector, lowers it by $5.29$, $5.79$ and $9.18$. The ratio
matters more than the ordering: an expansion history fixed by
Eq.~\eqref{eq:Lc} and the field equations recovers $57\%$ to $71\%$ of a
reduction that the parametrization obtains by fitting $w(a)$ to the same
data.

Most of that improvement comes from the low-redshift distances. Relative
to \LCDM, BDE-BDM lowers the DESI BAO $\chi^2$ by $2.95$ to $3.67$ and
the SNe $\chi^2$ by $2.53$ to $4.07$, with a further $0.74$ to $1.14$
from the low-$\ell$ temperature likelihood. Against \wowa\ the BAO
comparison is closer than the totals suggest: BDE-BDM reaches the lower
BAO $\chi^2$ for PantheonPlus ($9.93$ against $10.24$) and DES-Dovekie
($10.12$ against $10.42$), and trails only for Union3.1 ($10.30$ against
$8.79$). One component runs the other way. In the CamSpec high-$\ell$
likelihood \wowa\ improves on \LCDM\ by $2.63$ to $3.24$ while BDE-BDM
worsens it by $0.64$ to $3.00$, which accounts for almost the whole
difference between the two models' totals. This is a fractional change
of order $3\times10^{-4}$ in a likelihood contributing
$\chi^2\simeq10\,546$, confined to the damping tail, and it originates
in the extra radiation \Next\ present before condensation, which cannot
be tuned away.

The complexity penalty changes this picture. The effective parameter
count is
$p_D=13.06$--$14.49$ for BDE-BDM, $14.64$--$15.80$ for \wowa\ and
$13.58$--$15.35$ for \LCDM, in each case close to the number of
well-constrained parameters. Through
Eq.~\eqref{eq:dic_chi2_identity} these give
$\Delta\mathrm{DIC}=-8.28$, $-7.52$ and $-7.47$ for BDE-BDM, all in the
strong range and differing by only $0.81$ between compilations. The
\wowa\ values, $-6.84$, $-11.63$ and $-9.76$, are strong for
PantheonPlus and DES-Dovekie, decisive for Union3.1, and span $4.8$. The
contrast says more about the data than about the fits: the unified model
cannot reshape $w(z)$ to absorb a systematic difference between
compilations, so it responds only to the trend they share.

The evidence asks not how well a model fits at its best point but how
well it fits on average over the prior it was given, so a model that
buys its improvement with a large prior volume is charged for it.
Writing $\ln\mathcal{Z}=\langle\ln\mathcal{L}\rangle_{\mathcal{P}}
-\mathcal{D}_{\rm KL}$, with $\mathcal{D}_{\rm KL}$ the
Kullback--Leibler divergence of the posterior from the prior; the Occam
penalty is the second term. For \wowa\ the two nearly cancel: the
likelihood gain is real, but it is obtained by exploring
$w_0\in[-3,1]$ and $w_a\in[-3,2]$, of which the data select a small
part, and the Bayes factors of Table~\ref{tab:dic_full} leave no
preference for \wowa\ despite $\Delta\chi^2$ between $-9$ and $-13$. The
same cancellation is reported by independent analyses of these
data~\cite{Ong2025,Dinda2026}. BDE-BDM has no additional dark-energy
parameter to pay for: its likelihood gain of $3.3$ to $4.2$ meets a
compression difference of only $-0.66$ to $-0.79$ and survives as
moderate evidence over \LCDM\ for every compilation.

Tables~\ref{tab:dic_full} and~\ref{tab:chi2_all} also include BDE-CDM,
the parent model of Ref.~\cite{MacorraLozano2025}, in which the condensed
dark group forms bound dark energy alone and the dark matter is cold,
constrained with the same data. It improves on \LCDM\ by
$\Delta\chi^2=-1.98$, $-1.07$ and $-4.70$, again through the BAO and
supernova terms, but by less than BDE-BDM in every combination, and it
fits the CMB-lensing likelihood worse than \LCDM\ by $0.7$--$1.9$.
Its $\Delta\mathrm{DIC}$ of $-1.77$, $-2.03$ and $-1.56$ lies below the threshold of
$2$ for PantheonPlus and DES-Dovekie and just reaches the positive range
for Union3.1, and its Bayes factors over \LCDM, $+1.35$, $+2.26$ and
$+2.08$, are weak evidence. Against the parametrization it fares
better, $\ln B_{\mathrm{BDE\text{-}CDM},\wowa}=+3.46$, $+2.52$ and
$+2.64$, moderate evidence for every compilation. BDE-BDM exceeds
it by $\ln B=+2.74$, $+1.86$ and $+2.74$ and lowers $\chi^2$ relative to it
by $3.3$--$4.7$, led by the BAO.

Two further models complete the comparison. \LBDM\ and \wowaBDM\ pair
bound dark matter with a cosmological constant and with the CPL
parametrization, so that each dark-energy sector appears with both cold
and bound dark matter; with no condensing dark-energy sector,
Eq.~\eqref{eq:acLc} no longer fixes \ac, which is sampled with the priors
of Table~\ref{tab:priors}. \wowaBDM\ attains the best fit of
the six models, $\Delta\chi^2=-10.10$, $-11.83$ and $-13.42$, improving on
\wowa\ by $0.5$--$1.6$, and its $\Delta\mathrm{DIC}$ of $-6.89$, $-10.35$
and $-10.19$ is strong to decisive. With four parameters beyond those of
\LCDM\ it also pays the largest Occam penalty, and its Bayes factors over
\LCDM, $-0.82$, $+1.05$ and $+0.62$, express no preference; BDE-BDM is
favored over it by $\ln B=+4.92$, $+3.07$ and $+4.21$. \LBDM\ reaches
$\ln B=+2.49$, $+2.51$ and $+2.06$ over \LCDM. The preference for bound
dark energy does not depend on the dark matter: it is favored over a
cosmological constant by $+1.35$ to $+2.26$ with cold dark matter
(BDE-CDM against \LCDM) and by $+1.60$ to $+2.76$ with bound dark matter
(BDE-BDM against \LBDM), and over CPL by $+3.07$ to $+4.92$ with bound
dark matter. Bound dark matter improves the fit where the dark energy is predicted, lowering $\chi^2$ by $4.3$--$4.8$ from \LCDM\ to \LBDM\ and by
$3.3$--$4.7$ from BDE-CDM to BDE-BDM, in both cases led by the BAO, but by
only $0.5$--$1.6$ from \wowa\ to \wowaBDM, whose free parameters already
absorb the BAO pull. BDE-BDM has the largest evidence of the six.

Set directly against each other rather than against \LCDM, \wowa\ and
BDE-BDM are close. The difference
$\mathrm{DIC}_{\rm BDE\text{-}BDM}-\mathrm{DIC}_{\wowa}$ is $-1.44$,
$+4.12$ and $+2.29$: marginally ahead with PantheonPlus, behind with the
other two, decisive nowhere. Little should be read into this. An
absolute scale exists only between nested classes, and no continuous
limit of BDE-BDM returns either alternative, so following
Ref.~\cite{Lodha2025} for its own non-nested classes we quote
$\Delta\mathrm{DIC}$ against \LCDM\ only; and $p_D$ carries an
uncertainty of order unity, moving between $13.58$ and $15.35$ across
the \LCDM\ entries of Table~\ref{tab:dic_full} alone. On the DIC, the
two dynamical models stand on a par; the Bayes factors, which do charge
for prior volume, separate them. In $\Delta\chi^2$, the unified model captures the
greater part of what \wowa\ achieves over \LCDM\ with no dark-energy
parameter beyond those of \LCDM, while also predicting a small-scale
suppression on which the parametrization is silent. In
$\Delta\mathrm{DIC}$ it lies in the strong regime against \LCDM\ for
every supernova sample.

\section{Discussion}
\label{sec:discussion}

Unlike the \wowa\ model, the BDE-BDM model predicts the shape of
$w(z)$ with no dark-energy parameter beyond \Lc, and the fit determines it to
$w_0=-0.9262\pm0.0005$. Improved BAO and SNe data over
$0.2\lesssim z\lesssim2$ therefore test the model without additional
freedom. The redshift-resolved constraints of Figs.~\ref{fig:apgrid} and~\ref{fig:dvgrid} show where
that test is sharpest, the separation between models growing with
redshift in both the geometrical and the growth observables.

 The free-streaming cutoff of Sec.~\ref{sec:pkresult} falls in the interval covered by the
Lyman-$\alpha$ forest and by weak lensing at small angular separations,
the scales the component was originally introduced to
address~\cite{Macorra2010,Mastache2020}. This is
the most immediate test of the model, because the present data leave
\vc\ largely undetermined while the observable consequences of the
velocity are confined to precisely these scales. However, future
observations at smaller scales can restrict \vc\ further. What the cutoff separates is not only the unified model from \LCDM\ but
the model from its own cold limit, where the BDE-driven enhancement of small-scale power found for BDE-CDM~\cite{MacorraLozano2025} reappears in its place; whether the
deviation is positive or negative is thus itself a measurement of \vc.
Constraints from compressed Lyman-$\alpha$ summary statistics bound the
model on one side alone, since the observable saturates once the cutoff
leaves the range of the forest; a dedicated analysis with the full flux power
spectrum, and a comparison of the predicted halo mass function with
satellite counts, are left for future work.

Finally, the unified model contains two informative limits that were
studied before it was assembled. For $\fbde=1$ the entire dark group
condenses into the meson; with the dark matter supplied by a separate
cold component, the model reduces to
BDE-CDM~\cite{Almaraz2019,MacorraLozano2025}. Conversely, replacing the
meson by a cosmological constant and freeing the transition epoch
isolates the bound-dark-matter component~\cite{Macorra2010,Mastache2020},
the \LBDM\ model of Sec.~\ref{sec:comparison}.
The unified BDE-BDM model ties the two sides together: one gauge sector,
one transition, and one energy-conserving partition, which make its
predictions correspondingly more strongly.

\section{Conclusions}
\label{sec:conclusions}

We have set out the theory of the unified Bound Dark Energy--Bound Dark
Matter model and its confrontation with current data. What distinguishes it from
phenomenological descriptions of an evolving dark-energy density is that
the dark energy and the dark matter are not postulated
separately: both emerge from one confining dark gauge sector, by the
non-perturbative mechanism that binds quarks into hadrons. Three of its
ingredients are consequences rather than choices---the condensation
scale of gauge-coupling unification, Eq.~\eqref{eq:Lc}; the transition
epoch of consistency with the radiation background,
Eq.~\eqref{eq:acLc}; and the division of the condensing energy of energy
conservation, Eq.~\eqref{eq:fbde}. Only the bound velocity \vc\ is
fitted, and it belongs to the dark-matter sector, so that the model
samples one parameter more than \LCDM\ and one fewer than \wowa.

Sampling \Lc\ within the prior set by the theoretical prediction places
the condensation scale at $\Lc=43.67\pm0.13$~eV, in the interior of the
range $34^{+16}_{-11}$~eV that unification-scale physics gives, and
stable across the three supernova compilations. Because the prior
already encodes the prediction, this is a consistency result and not an
independent confirmation of it; what the data add is that nothing in the
fit pushes \Lc\ towards either edge of the window. The transition epoch
that follows, $\ac=(2.505\pm0.008)\times10^{-6}$, satisfies the bound
from the standalone bound-dark-matter
analyses~\cite{Macorra2010,Mastache2020} in the cold limit, the regime to
which small-scale data are expected to drive \vc, so the epoch at which
the dark energy must condense is early enough for the bound dark matter
to form,
a check that ties the two halves of the model together rather than
testing either one alone. From the same condensation the equation of
state follows without further input, thawing to
$w_0=-0.9262\pm0.0005$ today and remaining above $-1$ throughout, which
keeps the model clear of the instabilities that attend a phantom
crossing. Its narrow uncertainty measures how tightly the theory binds
$w(z)$ to \Lc, not how sharply the data discriminate, and it too is
stable across the compilations.

The same transition sets the amplitude of structure, lowering \sig\ to
$0.782\pm0.008$ for the baseline data with Union3.1 supernovae. The
reduction comes from two sources that differ in how firmly each is
established. About $0.012$ of it is produced by the transient extra
radiation and the dark-energy dynamics and survives for any \vc; the
remainder comes from free streaming and depends on the velocity, as does
the whole of the reduction in \Seight, whose cold-limit value coincides
with that of \LCDM. The velocity itself is only weakly constrained,
$\vc=0.463\pm0.171$, matching the earlier finding that background, SNe
and CMB data leave it open~\cite{Mastache2020}. However, future
observations at smaller scales can restrict \vc\ further. Those scales
also separate the model from its own cold limit, since with cold dark
matter the spectrum is enhanced by up to $25\%$ near
$k\approx4.3\,\mathrm{Mpc}^{-1}$ rather than suppressed, and the sign of
the small-scale deviation reverses.

None of the dark-energy results depend on that velocity.
Section~\ref{sec:coldlimit} shows that restricting the chains to the
cold limit moves the condensation scale, the transition epoch and the
expansion rate by less than half a standard deviation and leaves the
predicted $w(z)$ unchanged, so these results do not wait on the
small-scale measurements that will eventually fix \vc. Set against \LCDM, the deviance information criterion places BDE-BDM in
the strong range for all three compilations,
$\Delta\mathrm{DIC}=-8.28$, $-7.52$ and $-7.47$, a stability across
samples that the \wowa\ values, $-6.84$ to $-11.63$, do not have. The
Bayesian evidence draws the two descriptions further apart. \wowa\ gains
$\Delta\chi^2=-9.3$ to $-12.9$ and then returns it to the prior volume
of its two extra parameters, leaving $\ln B=-2.11$, $-0.26$ and $-0.56$;
BDE-BDM has no additional dark-energy parameter to pay for and keeps
$\ln B=+4.09$, $+4.12$ and $+4.83$, moderate on the Jeffreys scale. The DIC comparison between
BDE-BDM and \wowa\ carries no absolute meaning, the two models being
non-nested. Extending the comparison to \LBDM\ and \wowaBDM\ shows that
bound dark energy is preferred over a cosmological constant with either
cold or bound dark matter, and over the CPL parametrization when the dark
matter is bound. 

The future surveys now underway and in preparation will deliver all three at the precision required. Should the model survive them, the cosmological constant will have been replaced not by a fitted function of redshift but by a condensate, and
the dark energy and the dark matter will trace back to a single
event in the history of one gauge sector.

\begin{acknowledgments}
We acknowledge financial support from PAPIIT-DGAPA-UNAM (101124),
SECIHTI project CBF-2025-G-1299, and the SECIHTI scholarship grant for
Jose Lozano.
\end{acknowledgments}

%

\end{document}